# A Spectral Analysis of Business Cycle Patterns in UK Sectoral Output

Peijie Wang*
Business School
University of Hull
Hull HU6 7RX
Email : p.wang@hull.ac.uk

and

Trefor Jones
Manchester Business School
University of Manchester
Booth Street West
Manchester M15 6PB

**Abstract** - This paper studies business cycle patterns in UK sectoral output. It analyzes the distinction between white noise processes and their non-white noise counterparts in the frequency domain and further examines the associated features and patterns for the process where white noise conditions are violated. The characteristics of these sectors, arising from their institutional features that may influence business cycles behavior and patterns, are discussed. The study then investigates the output of UK GDP sectors empirically, revealing their similarities and differences in their business cycle patterns.

* corresponding author

## 1. Introduction

This paper examines business cycle patterns in UK GDP sectors in the frequency domain. It analyses the spectra of sectoral output and focuses on the way empirical spectra behave across GDP sectors. This approach offers a different means of research through inspecting the degree to which the time series deviates from a white noise process or its integral, a pure random walk, an indication of the relative importance of the cycle in the time series. Moreover, the pattern in the spectrum explains how the time series deviates from a pure random walk. This is helpful as cycles themselves differ from one type to another. Therefore, not only the weight of cycles in the time series, but also the behavior and compositions of cycles, may be made known through such scrutiny. Analysis in the frequency domain, or spectral analysis, is particularly helpful in the study of the relative contribution of each of the components in the time series variable, which establishes the overall pattern and behavior of the variable. Therefore, while generating no more information than we have in the time domain, the approach in the frequency domain may present a fuller picture of business cycle fluctuations, because it uses and processes the information in a more effective way for this type of investigation.

It is to a large extent accepted that most economic time series are non-stationary in their levels. However, it is not enough to decide whether an economic time series is stationary or non-stationary. Even if a time series is non-stationary in its level, its behavior can be quite different. Some economic and financial time series may consist of both stationary and non-stationary components and, consequently, there might be mean-reverting tendency. Therefore, the appropriate question to be asked and answered is not whether a time series is stationary or not. Instead, the right questions are: (a), as rightly pointed out by Campbell and Mankiw (1987a,b) and Cochrane (1988), how large is the random walk component in the time series; and (b), beyond that and investigated in this study, the patterns of departure from a pure random walk or its difference, a white noise process, in the time series.

The present study extends the work of Campbell and Mankiw (1987a,b) from the time domain to the frequency domain. It extends the work by Cochrane (1988) in the sense that Cochrane's measure in his 1988 paper "How big is the random walk in GDP?" could in fact be based on spectral analysis in the frequency domain too, but it was one specific point on the spectrum - the zero frequency point. This study goes beyond the question of how big the random walk component is in an economic time series by further examining the associated features and patterns for the process where white noise conditions are violated, pertinent to economic activity. The study attempts to elaborate on the institutional background of various sectors and the economic explanations for a particular business cycle pattern to be associated with a particular sector, and then examines how sectors behave differently and similarly.

It is of fundamental relevance to study the distinction between white noise processes and their non-white noise counterparts of general stationary processes of I(0), or between their respective integrals, pure random walks and general non-stationary processes of I(1). Firstly, it is of prime importance in finance at least for the sake of a weakly efficient financial market. Secondly, in general economic research terms, patterns of violation of white noise conditions reveal the characteristics or behavior of the process under investigation, which is of considerable empirical relevance to economic policy and business strategy, those related to business cycles in particular. First formal research of white noises and white noise conditions in the frequency domain can be attributed to Grenander and Rosenblatt (1953, 1957), which has been followed by a few of later studies, e.g., Priestley (1981, 1996). In his investigation of the departure of economic time series from a pure random walk process, Cochrane (1988) adopts an approach that appears to be in the time domain but indeed is a special case in the frequency domain at the zero frequency point. It is not exaggerated to claim that the first studies of business cycles were in the frequency domain, in as early as the first half of the 20th century, when the notion of business cycles started to attract attention from economists and governments alike in their search for an understanding of the patterns in economic activity and a possible therapy for mitigating the damage caused by severe economic downturn. Indeed, Schumpeter's (1939) long waves are low frequency cycles in the frequency domain. Although most empirical studies since then have been in the time domain, the significance of the frequency domain method in business cycle studies has been gradually acknowledged in the last decade. For example, Baxter and King (1999) develop several approximate band-pass filters in the frequency domain and apply these filters to the measurement of business cycles. The research by King and Watson (1996) on the relationship between money, prices, interest rates and business cycles is also in the frequency domain. More recently, A'Hearn and Woitek (2001) resort to the frequency domain method of spectral analysis to examine business cycles in 13 countries, using annual historical industrial output (industrial production) data from around 1865 to 1913. It can be observed that applications of the frequency domain method in economics and finance have been scarce and the progress has been slow. The present paper attempts to contribute to the development of this important analytical approach in general and its application in business cycle studies in particular.

The present study raises and attempts to answer these questions: (a) Does sectoral output follow a pure random walk? (b) If not, what patterns do they exhibit? (c) What are the economic explanations and the institutional background for a particular business cycle pattern to be associated with a particular sector? and (d) How do sectors behave differently and similarly? Since our approach developed and adopted in this study possesses noteworthy advantages, our analysis and empirical results lead to a better understanding of business cycle behavior. The rest of the paper is organized as follows. Section 2 examines the statistical distributions of time series, in particular, near white noise processes in the frequency domain, presenting and discussing patterns of violation of white noise conditions. Section 3 provides a brief inspection and discussion of the institutional features of the sectors that are concerned with the extent to which a sector is subject to the influence of a range of specific factors: regulatory requirements and government policy, foreign competition, dependence on demand, and the role of innovations and supply in the creation of demand. Section 4 carries out empirical investigations of business cycle patterns in UK sectoral output, reporting empirical findings and discussing their implications in relation to those factors of influence. Finally, Section 5 concludes.

## 2. Frequency domain analysis of the patterns of violation of white noise conditions

Spectral analysis, or studies in the frequency domain, is one of the unconventional subjects in time series econometrics. Analysis in the frequency domain does not bring in new or additional information, it is simply an alternative method with which information is observed, processed and abstracted. Analysis in the frequency domain is particularly useful to the examination of cyclical movements in prices, returns, and output. As the name suggests, it models and investigates time series variables according to their frequency components, therefore, has the advantages to distinguish patterns featuring e.g., shorter and longer cycles, and to reveal characteristics ascribing to higher and lower frequencies. This is sometimes helpful. Depending on the characteristics of the issues, analysis in one domain may be more powerful than in the other. For example, cycles are better and more explicitly observed and represented in the frequency domain. It is worthwhile pointing out that correlations in the time domain and cross spectra in the frequency domain deal with the relationship between two time series from different perspectives and have defined links.

2.1. Statistical distributions in the frequency domain of near white noise processes

Let us assume a time series $X(t)$ possesses the usual properties that it is stationary, is continuous in mean squares, and has higher moments up to the fourth moment, then the spectrum of the process, or the spectral distribution function, exists with the following relationships:

$$\begin{aligned} f(\omega) &= \frac{1}{2\pi N}\sum_{\tau=-N}^{N} R(\tau)e^{-i\tau\omega} = \frac{1}{2\pi N}\sum_{\tau=-N}^{N} R(\tau)\cos(\tau\omega) \\ &= \frac{\sigma_X^2}{2\pi} + \frac{1}{\pi}\sum_{\tau=1}^{N} R(\tau)\cos(\tau\omega) = \frac{C_0}{2\pi N} + \frac{1}{\pi N}\sum_{\tau=1}^{N-1} C_\tau \cos(\tau\omega) \end{aligned} \tag{1}$$

$$R(\tau) = \sigma_x^2 \int_{-\pi}^{\pi} e^{i\tau\omega} dF(\omega) \tag{2}$$

where $C_\tau = \sum_{t=1}^{N-\tau} X_t X_{t+\tau}$, $C_0 = \sum_{t=1}^{N} X_t X_t = N\sigma_X^2$, and $F(\omega) = \int_0^{\omega} f(\omega)d\omega$ is the integral spectrum of the process.

For a pure white noise process, $C_0$ obeys a $\chi^2$ distribution with $E\{C_0\}=N$, $Var\{C_0\}=2N$; and $C_\tau$ obey normal distributions with $E\{C_\tau\}=0$, $Var\{C_\tau\}=N$, for $\tau=1,...N-1$. $C_0$ is stochastic in both domains but previous research only considers it to be stochastic in the time domain but a constant in the frequency domain, from which this study differs. In the following, we show how a white noise process is distributed in the frequency domain, and the conditions on which a particular process can be accepted as a white noise process. We call such a process near white noise processes in contrast to a pure theoretical white noise. It can be shown that:

$$\lim_{N\to\infty} P\left\{\max_{0\le\omega\le\pi} N^{1/2}\left|F(\omega)-\frac{\omega}{2\pi}\right| \le \alpha\right\} = P\left\{\max_{0\le\omega\le\pi}|\xi(\omega)| \le \alpha\right\} \tag{3}$$

where $\xi(\omega)$ is a Gaussian process with:

$$P\{\xi(0)=0\}=1 \tag{4a}$$

$$P\{\xi(\pi)=0\}=1 \tag{4b}$$

$$E\{\xi(\omega)\} = 0, \quad 0 \le \omega \le \pi \tag{4c}$$

$$E\{\xi(\nu)\xi(\omega)\} = \frac{3\nu(\pi - \omega)}{4\pi^2}, \quad 0 \le \nu < \omega \le \pi \tag{4d}$$

$$E\{[\xi(\omega)]^2\} = \frac{3\omega(\pi - \omega)}{4\pi^2}, \quad 0 \le \omega \le \pi \tag{4e}$$

See Appendix for proofs.

There are two major conclusions from the above result: (a) a Gaussian process in the time domain with its variance being constant at every time point is also Gaussian in the frequency domain; but (b), its variance in the frequency domain is a function of $\omega$, it peaks at the point $\omega$ = $\pi/2$ and is zero at the two ends $\omega$ = 0 and $\omega$ = $\pi$. The property in (b) is in contrast to its time domain counterpart.

## 2.2. Patterns of violation of white noise conditions

This part discusses and abstracts typical patterns in time series where white noise conditions are violated. Behavior of a particular process will be examined, in accordance with its frequency domain characteristics, which is of more empirical relevance. From Equations (3) and (4), three propositions can be developed with regard to patters of violation of white noise conditions, setting against the benchmark of a white noise process.

**Proposition 1.** Lower frequency components stochastically dominate higher frequency components in the frequency range ($\varpi_1$, $\varpi_2$) if $\xi(\omega) > 0,\ \varpi_1 < \omega < \varpi_2$. Lower frequency components stochastically *consistently* dominate higher frequency components if $\xi(\omega) > 0,\ 0 < \omega < \pi$, and the time series is said to possess the features of the compounding effect.

By definition and according to the statement of equation (3), $\xi(\omega)$ is the difference between the integral of the process under examination and the integral of a pure white noise process, being scaled by $N$, when $N \to \infty$, i.e., $\xi(\omega) = \underset{N\to\infty}{Lim} N^{\frac{1}{2}} \int_0^{\omega} \left( I_p(\omega) - \frac{1}{2\pi} \right) d\omega$. Figure 1 shows the features of such stochastic processes. The top panel of the figure is the time domain response to a unit size shock of a time series with compounding features, against a random walk response. The dashed line indicates the evolution path of the time series if no shocks have ever occurred. The middle panel is a typical spectrum for such time series, and the bottom panel is the $\xi(\omega)$ statistic for such time series. The spectrum in Figure 1 is a monotonous decrease function of $\omega$, with $\xi(\omega) > 0,\ 0 < \omega < \pi$, and $\xi_{\omega}^{''}(\omega) < 0,\ 0 < \omega < \pi$, where $\xi_{\omega}^{''}(\omega)$ is the second order derivative of $\xi(\omega)$ with respect to $\omega$.[1] Stochastically consistent dominance has a looser requirement than a spectrum of monotonous function.

**{Figure 1 about here}**

**Proposition 2.** Higher frequency components stochastically dominate lower frequency components in the frequency range ($\varpi_1$, $\varpi_2$) if $\xi(\omega) < 0,\ \varpi_1 < \omega < \varpi_2$. Higher frequency components stochastically *consistently* dominate lower frequency components if $\xi(\omega) < 0,\ 0 < \omega < \pi$, and the time series is said to possess mean-reverting tendencies.

Figure 2 shows the features of such stochastic processes. The top panel of the figure is the time domain response to a unit size shock of a time series with mean-reverting tendencies, against a random walk response. The dashed line indicates the evolution path when there are no shocks to the time series. The middle panel is a typical spectrum for such time series, and the bottom panel is the $\xi(\omega)$ statistic for such time series. The spectrum in Figure 2 is a monotonous increase function of $\omega$, with $\xi(\omega) < 0,\ 0 < \omega < \pi$, and $\xi_{\omega}^{''}(\omega) > 0,\ 0 < \omega < \pi$. Stochastically consistent dominance has a looser requirement than a spectrum of monotonous function. Relevant discussions for Proposition 1 apply.

[1] $\xi_{\omega}^{'}(\omega) = N^{\frac{1}{2}}\left[I_p(\omega) - 1/2\pi\right], \xi_{\omega}^{''}(\omega) = N^{\frac{1}{2}} I_p^{'}(\omega)$.

**{Figure 2 about here}**

**Proposition 3.** Higher (lower) frequency components do not stochastically *consistently* dominate lower (higher) frequency components if there exist sub-sets of frequencies $\omega^{+}$, $\omega^{-}$ and $\omega^{0}$ that $\xi(\omega) > 0, \omega \in \omega^{+}$, $\xi(\omega) < 0, \omega \in \omega^{-}$ and $\xi(\omega) = 0, \omega \in \omega^{0}$; and the time series is said to possess the features of mixed complexity.

Figure 3 demonstrates the features of such stochastic processes. Relevant discussions for Proposition 1 apply. Figure 3(a) shows a case where there are more powers in the medium range frequencies, while Figure 3(b) shows a case where there are more powers in the low and high frequencies. The top panel of the figures is the time domain response to a unit size shock of a time series with the features of mixed complexity, against a random walk response. The dashed line indicates the evolution path when there are no shocks to the time series. The middle panel is typical spectra for such time series, and the bottom panel is the $\xi(\omega)$ statistics for such time series.

**{Figure 3 about here}**

## 3. Institutional features of the sectors and business cycle patterns

Prior to formal empirical analysis of the response patterns of sectoral output in business cycles, a brief inspection and discussion of the institutional features of the sectors would be helpful. These features are concerned with the degree and/or extent to which a sector is subject to the influence of a range of specific factors. Only those institutional features that are most relevant to a sector's distinct response patterns in business cycles are considered: (a) regulatory requirements and government policy; (b) intensity of foreign competition; (c) dependence on demand; and (d) the role of innovations, and subsequently supply, in the creation of demand in new forms or shapes. A regulated industry's output and prices are not driven entirely by market forces, so its business cycle patterns can be different from those of unregulated industries. Government policy includes the impact of both domestic and foreign government policy, as well as the common policy of groups of nations, such as the EU and OPEC. Similar to the regulative effect, the output and prices of a sector that is subject to government policy to a large extent would behave rather differently from those sectors that are less influenced by government policy. In general, a sector with a higher degree of influence by regulation and government policy would show relatively more persistent response patterns in business cycles, other things being equal. Foreign competition in this research considers the impact of foreign competition on the domestic output of a sector and is not concerned with ownership. For example, it is not taken into account whether the energy sector is 60 percent owned by foreign multinationals or 100% owned by domestic companies; the criterion is the proportion of the final product that is produced domestically, instead of being imported, or export in the case of competition abroad. A country may virtually have no manufacturing while its residents consume the same amount or more of manufactured goods as the residents in other countries. However, this does not apply to some sectors, e.g., construction, which must employ a proportional workforce to produce proportional output domestically, at least in the case of the UK. A sector subject to intensive foreign competition would endure incessant challenges from abroad and, depending on its competitive advantages or disadvantages, would experience steady and continual shrinkage or expansion. Either way, the consequence is that its response patterns in business cycles are more persistent or with compounding effects. A sector that is largely demand led show less persistent, mean-reverting, response patterns in business cycles as it is subject to demand shocks that are temporal. Finally, a sector in which individual companies' survival and growth are featured by innovations and the reliance on innovations to create and generate demand in new forms or shapes is subject to supply shocks and demonstrates relatively more persistent response patterns in business cycles.

Then the seven main sectors used in the study are: Agriculture, Forestry and Fishing (A&B); Manufacturing (D); Electricity, Gas and Water Supply (E); Construction (F); Distribution, Hotels, Catering and Repairs (G&H); Transport, Storage and Communication (I); and Services (J-Q, including business services and finance, and government and other services). The Mining and Quarrying sector (C) is excluded, as its weight in UK GDP is minimal and has being declining over decades; and more importantly, its change has been mainly influenced by unconventional economic forces and other factors. The Services sector is examined as a whole as well as in two parts of Business Services and Finance (J&K) and Government and Other Services (L-Q), since the attributes and features of these two types of services are rather different and, consequently, may possess different response patterns in business cycle fluctuations. However, the two disaggregate services series only came into existence in the first quarter of 1983, instead of the first quarter of 1955 for the seven main sectors. For comparison purposes, the aggregate Services sector is also investigated for the period starting in the first quarter of 1983, in addition to the period starting in the first quarter of 1955.

**{Table 1 about here}**

We summarize the features of the sectors in Table 1, in accordance with the above analysis. For example, sector A&B, Agriculture, Forestry and Fishing, is highly influenced by government policy, e.g., the French government's subsidy policy in agriculture and the EU's common agriculture policy. It endures severe foreign competition and is largely demand led; new product or flavor, e.g., GM food, does not have a significant impact on consumption and new demand. From the perspectives of government policy and foreign competition, the sector is expected to show higher persistence in its business cycle patterns; however, from its perspectives of demand and supply, the sector would exhibit less persistent, or mean-reverting, patterns. Taking all these features into consideration, the business cycle patterns of the Agriculture, Forestry and Fishing sector are a matter of empirical investigation. In general, sector D, Manufacturing, is not subject to strict regulatory requirements and government policy. However, its exposure to foreign competition is high. Demand is critical for the sector's output, so are innovations and R&D, the supply side factors, to maintain and generate new demand and compete with imported foreign manufactured goods. Overall, the Manufacturing sector is expected to exhibit higher persistence in its business cycle patterns. Sector E, Electricity, Gas and Water Supply, or the energy sector, is a regulated industry. Its exposure to foreign competition is low, though the sector is foreign owned to a large extent, its final products and supply to consumers are mainly domestically based. It is a typical demand led sector, with the role of supply side factors in establishing output levels being minimal. From the perspective of regulation, the sector is expected to show higher persistence or compounding effects in its business cycle patterns; however, from its perspective of foreign competition and that of demand and supply factors, the sector would exhibit less persistent, or mean-reverting, patterns. Consequently and overall, the energy sector would exhibit less persistent, random walks with mean-reverting tendency or mixed complicity, patterns in business cycles, as the compounding effect would be largely overpowered by mean-reverting tendencies. Except that regulatory requirements are low, sector F, Construction, possesses the features similar to the energy sector, and is expected to show less persistent, or random walks with mean-reverting, response patterns in business cycles. Sector G&H, Distribution, Hotels, Catering and Repairs, is also typical demand led with low regulatory requirements. Moreover, demand for goods and services in parts of this sector is highly variable due to the nature of such consumptions, and the variability in demand is as durable as business cycles. It faces medium to high degrees of foreign competition in a subtler way, with the hotel industry being the most explicit. From the perspective of foreign competition, the sector would exhibit higher persistence or compounding effects in its business cycle patterns; but the high variability in demand and the high durability of the variability in demand indicate that, though a demand led industry, the effect of demand/supply factors is not simply mean-reverting, but can be rather persistent or mixed. Therefore, this sector is expected to exhibit some weak compounding effect with mixed complicity in its response patterns in business cycles. Sector I, Transport, Storage and Communication, also falls into the reign of regulation. It relies on innovations, R&D and investment in infrastructure, the supply side factors, heavily. Nevertheless, its exposure to foreign competition, in the sense of non-domestically produced final products and services, is low. Overall, this sector is expected to exhibit higher persistence in its business cycle patterns due to a high degree of regulatory requirements and the contribution of the supply side factors. Sector J-Q, the Services sector, is divided into Business Services and Finance, J&K, and Government and Other Services, L-Q. The Business Services and Finance sector, the only major sector in which the UK enjoys certain comparative advantages, is subject to severe foreign competition and witnessed some decline in the last two decades, following the steady decline in the Manufacturing sector. Supply side factors, such as innovations and new methods of doing business, are as critical as demand to the survival and growth of the companies in this sector. Consequently, the sector is expected to show higher persistence in its response patterns in business cycles. Lastly, sector L-Q, Government and Other Services (social services and other non-profitable services), is expected to be rather different from most of the other sectors. It is subject to least foreign competition, but is directly linked to government policy. Nevertheless, the way in which the sector is affected by government policy is different. In addition, the sector is supply driven, not demand led, but the supply driven mechanism is rather different from that in the other sectors, which is not mainly to do with technology shocks, such as innovations and R&D, but with government policy. It is not a steady proportion of GDP either as government spending does not necessarily increase when GDP increases or government spending may increase when GDP falls. So the behavior of the sector does not mirror that of GDP. Despite all these complications, one thing is sure: the sector experiences relatively infrequent shocks than the other sectors. As a result, its output series could look fairly stable and, consequently, stationary in its appearance. Therefore, this sector may be expected to exhibit high mean-reverting tendencies in its response patterns in business cycles.

## 4. Data, empirical study and discussions

### 4.1. Data sets and summary statistics

The data sets used in this study are UK aggregate GDP and output in seven main GDP sectors, the institutional features of which have been examined in the previous section. The data sets form the aggregate GDP and seven main GDP sectors start in the first quarter, 1955, end in the first quarter, 2002, and are seasonally adjusted at the 1995 constant price. The data sets for the two sub-sectors within the Services sector start from the first quarter in 1983.

**{Table 2 about here}**

Summary statistics of these sectors' output and GDP are provided in Table 2. Sector J&K, Business Services and Finance (from 1983), sector I, Transport, Storage and Communication, and sector E, Electricity, Gas and Water Supply, enjoy a greater than average growth rate, though the Business Services and Finance sector has experienced a decrease in its growth rate. One of the prominent casualties in the Business Services and Finance sector is the shift and outsourcing of work to India and other English speaking countries to conduct administrative work and insurance business, such as insurance claims. On the list of the companies have been Prudential, Norwich Union, and Goldman Sachs. The shift and outsourcing has offset the legendary success of exports in education and related services to a measurable extent, due to the same reason of the English language being one of the most commonly spoken languages in the world. The lowest growing sectors are A&B, Agriculture, Forestry and Fishing, and D, Manufacturing. The Manufacturing sector has also gone through a decline in its growth during this period, along with sector F, Construction. As being analyzed above, sector L-Q, Government and Other Services (from 1983), has the most smoothed growth, with its standard deviation in growth being the smallest and much smaller than that for all the other sectors. The most volatile sector is E, Electricity, Gas and Water Supply, followed by F, Construction, and A&B, Agriculture, Forestry and Fishing.

### 4.2. Empirical results and discussions

The estimated $\xi(\omega)$ statistics for sectoral output and GDP are plotted in the middle panel of Figures 4-13. We use confidence intervals to examine and assess the features of the process, which is easily perceptible. In addition, output series themselves are exhibited in the top panel and spectra are presented in the bottom panel of these figures. We examine the $\xi(\omega)$ statistic and inspect the associated patterns for the GDP sectors in relation to their institutional features reviewed earlier. Four sectors show the features of compounding effects to varied degrees. They are sector A&B, Agriculture, Forestry and Fishing; sector D, Manufacturing; sector I, Transport, Storage and Communication; and sector J&K, Business Services and Finance. This finding confirms our previous analysis of their institutional characteristics and the ways in which they are subject to the influence of a range of factors in relation to business cycle patterns. However, an empirical examination of these sectors' output data further renders us specific insights into the sectors. Among the four sectors, compounding effects in response to shocks are confirmed overwhelmingly in sector A&B and sector J&K in that the near white noise conditions are significantly violated – as shown in Figure 4 and Figure 11, $\xi(\omega)$ statistics are positive in the whole frequency range and the majority of $\xi(\omega)$ are substantially above the upper band of the 95% confidence interval. In the case of sector D, $\xi(\omega)$ are positive in the whole frequency range but only a small part of $\xi(\omega)$ are beyond the upper band of the 95% confidence interval, revealed by Figure 5. For sector I, it is observed in Figure 9 that most of $\xi(\omega)$ are positive and only a small part of $\xi(\omega)$ are beyond the upper band of the 95% confidence interval. So, compounding effects are not as strong in sector D and sector I as in sector A&B and sector J&K. Since these sectors possess the features of compounding effects in their response to shocks in business cycles, the consequence of good as well as bad events or incidents, policy related or technology based, would accumulate in the course to affect the performance of these sectors, with the Agriculture, Forestry and Fishing sector and the Business Services and Finance sector being hit the most.

As shown by Figure 10(b) and Figure 11, sector J-Q, the aggregate Services sector, possesses a similar business cycle pattern with sector J&K, Business Services and Finance, for the period staring from the first quarter of 1983. The aggregate Services sector is examined for this period for two reasons. Firstly, data for sector J&K, Business Services and Finance, and data for sector L-Q, Government and Other Services, are only available since 1983. Since the two sub-sectors are rather different, there is a need to study them individually. For comparison purposes, their aggregate is also examined in the same period. Secondly, as exhibited by Figure 10(a), there appears to be some problems in the data for the aggregate Services sector in the full period starting from the first quarter of 1955. There are regular cyclical oscillations observed in the spectrum, which is reflected in its $\xi(\omega)$ statistics as well. As a result, our analysis is based on the period starting from the first quarter of 1983 for the Services sector.

Sector E, Electricity, Gas and Water Supply, and sector F, Construction, demonstrate random walk like behavior – it is observed in Figure 6 and Figure 7 respectively that all the values of the $\xi(\omega)$ statistic are confined to the 95% confidence interval and the near white noise conditions are not violated. Between the two, the Construction sector exhibits a weak mean-reverting tendency, while the Electricity, Gas and Water Supply sector displays some weak features of mixed complexity, to a statistically insignificant degree. These findings reinforce our analysis of the two sectors' institutional features and confirm our early conjectural explanation that, between the two sectors, the Construction sector would display relatively less persistent response patterns in business cycles due to its lower regulatory requirements.

Sector G&H, Distribution, Hotels, Catering and Repairs, is associated with a mixed complicity response pattern in business cycles and exhibits some compounding effect to a certain extent also, as being demonstrated by Figure 8. Almost half of $\xi(\omega)$ statistics are positive and half of $\xi(\omega)$ statistics are negative, though only the positive part of $\xi(\omega)$ violate the near white noise conditions and are beyond the upper band of the 95% confidence interval. Some of the negative $\xi(\omega)$ statistics are close to, but yet to reach, the lower band of the 95% confidence interval. These findings fit into the institutional characteristics of the Distribution, Hotels, Catering and Repairs sector fairly appropriately.

Sector L-Q, Government and Other Services, as expected, exhibits a business cycle pattern rather different from that in all other sectors, revealed by Figure 12. It possesses mean-reverting tendencies to such an extent that is almost for a stationary time series. All the values of the $\xi(\omega)$ statistic are negative, most of them having violated the near white noise conditions and being below the lower band of the 95% confidence interval. We have observed earlier in Table 2 that the sector has the most smoothed growth, with its standard deviation in growth being much smaller than that for all the other sectors, mainly arising from the sector's characteristics of experiencing infrequent shocks in business cycles. Smoothed growth, or a small standard deviation in growth, does not necessarily mean a lower degree of persistence or close to being stationary. It is infrequent shocks that, to a large extent, contribute to the features demonstrated by the Government and Other Services sector.

**{Figures 4 - 13 about here}**

The behavior of the aggregate GDP must reflect the business cycle features demonstrated by GDP sectors that are dominated by persistent, sizeable compounding effects in their response to shocks in business cycles. It is observed in Figure 13 that the majority of $\xi(\omega)$ statistics are positive, with a few of them being beyond the upper band of the 95% confidence interval or having violated the near white noise conditions. Although the result from the analysis of the aggregate GDP makes known its business cycle response patterns and features, which match the outcome and conclusion of sectoral analysis, it is sectoral analysis and, in particular, the analysis of the institutional background and characteristics of the sectors, that reveals how different sectors behave differently in business cycles and why a specific sector exhibits a specific business cycle pattern, and lays theoretical cornerstones for GDP's overall business cycle features. This contribution makes the present study distinct in the literature.

## 5. Conclusion

In this paper business cycle patterns in UK GDP sectors have been examined. The paper has developed a frequency domain approach to analyzing the distinction between white noise processes and their non-white noise counterparts in the frequency domain. It has then examined the associated features and patterns for the process where white noise conditions are violated, and classified and summarized these features and patterns in a way that is of empirical relevance to business cycle research. The characteristics of GDP sectors, arising from their institutional features that may influence business cycles behavior, have been discussed, in conjunction with the classified frequency domain patterns. Empowered by this analytical approach, the present study has then investigated the output of UK GDP sectors empirically, revealing their features and contrasting their similarities and differences in their business cycle response patterns.

This empirical research differs from previous business cycle studies focusing on trend-cycle decompositions, making consequential contributions to the study of business cycle behavior. Firstly and technically, our method is to examine the way in which a time series deviates from a white noise process (or its integral, a pure random walk) - how the deviation takes specific forms, to identify and classify empirically relevant business cycle patterns and features. Secondly and theoretically, our analysis is centered on the extent to which a sector is subject to the influence of a range of specific factors, paying attention to the institutional background and features that are most relevant to a specific sector's distinct response patterns in business cycles, including regulatory requirements and government policy, intensity of foreign competition, dependence on demand, and the role of innovations and supply in the creation of demand in new forms. The effort to establish a close association between the summarized sectoral business cycle features and the classified frequency domain patterns makes the present study not only empirically relevant but also theoretically revealing.

Thirdly, it is concluded that it is the inspected aspects of the institutional features of the sectors that contribute to the specific business cycle response patterns of the sectors. Business cycle patterns do not merely demonstrate some stylized phenomena of time series data; they have profound economic and institutional foundations. In view of that, practices, such as business cycle forecasts focusing on data analysis alone, no matter how complicated and advanced they are, are of little help. Research that scrutinizes the influence on the sectors of specific factors helps link sectoral business cycle patterns, arising from their institutional characteristics, to their time series behavior, and tells a fundamental story about business cycle evolution and development.

Fourthly and finally, it is demonstrated and confirmed that UK output predominantly possesses persistent, sizeable compounding effects in its response to shocks in business cycles, as evidenced by the results for GDP and four out of seven main sectors. Business cycle response patterns and features of the aggregate GDP fit, as expected, into the outcome and conclusion of sectoral analysis. The results and findings, together with their reflective implications, help make fuller use of accessible knowledge and advance our understanding of the causes and progression of output fluctuations.

**Appendix: Proof of equation (3)**

The integral spectrum of the time series process, or the integral of the spectrum of equation (1), is:

$$F(\omega)=\int_0^{\omega} f(\omega)d\omega=\frac{v_n^*(\omega)}{2\pi N}+\sum_{\tau=1}^{N-1}\frac{C_\tau}{\pi N}\frac{\sin\tau\omega}{\tau} \qquad \text{(A1)}$$

where $v_n^*(\omega)=\int_0^{\pi} C_0(\omega)d\omega$. Then:

$$N^{1/2}\left[F_p(\omega)-\frac{\omega}{2\pi}\right]=\frac{N^{1/2}}{2\pi}\left[\frac{v_n^*(\omega)}{N}-\omega\right]+\sum_{\tau=1}^{N-1}\frac{C_\tau}{\pi N^{1/2}}\frac{\sin\tau\omega}{\tau} \qquad \text{(A2)}$$

Previous studies[2] treat $C_0$ as a stochastic variable in the time domain, but a constant in the frequency domain irrespective of the value of $\omega$. There is a problem mainly concerning the boundary condition: $C_0$ can be greater or smaller than $N$ at point $\omega=\pi$, which does not guarantee $F(\pi)=$ ½ [$F(\pi)$ is half of the total power], the requirement that the power of the standardized time series is unity (the second term on the right hand side of equation (6) is zero at point $\omega=\pi$). We resort to the Kolmogorov-Smirnov theorem for a realistic representation of the distribution of $C_0$.

The distribution of the first term on the right hand side can be obtained by applying the Kolmogorov-Smirnov theorem. Define:

$$z_n(t)=N^{1/2}\left[\frac{v_n(t)}{N}-t\right],\quad 0\le t\le 1 \qquad \text{(A3)}$$

where $v_n(t)-v_n(s)$ is the number of successes in $N$ independent trials, with probability t – s of success in each trial; $P\{v_n(0)=0\}=1$, $P\{v_n(1)=N\}=1$; $E\{v_n(t)-v_n(s)\}=N(t-s)$, $E\{[v_n(t)-v_n(s)]^2\}=2N(t-s)[1-(t-s)]$, $0\le s<t\le 1$;

[2] e.g., Bartlett (1950), Grenander and Rosenblatt (1953), and Priestley (1996).

$E\{[v_n(t_1)-v_n(s_1)][v_n(t_2)-v_n(s_2)]\}=0$, $0\le s_1<t_1\le s_2<t_2\le 1$. Then, the m-variable distribution of the random variables $z_n(t_1),\ldots, z_n(t_m)$, $0\le t_1<\ldots<t_m\le 1$ is Gaussian, and

$$P\{z_n(0)=0\}=1,\ P\{z_n(1)=0\}=1 \tag{A4a}$$

$$E\{z_n(t)\}=0,\quad 0\le t\le 1 \tag{A4b}$$

$$E\{[z_n(t)-z_n(s)]^2\}=(t-s)[1-(t-s)],\quad 0\le s<t\le 1 \tag{A4c}$$

Now, let $\omega=\pi t$ and define:

$$z_n^*(\omega)=\frac{N^{1/2}}{2\pi}\left[\frac{v_n^*(\omega)}{N}-\omega\right] \tag{A5}$$

we have:

$$P\{v_n^*(0)=0\}=1,\ P\{v_n^*(\pi)=N\}=1 \tag{A6a}$$

$$E\{v_n^*(\omega)-v_n^*(\nu)\}=N(\omega-\nu),\quad 0\le\nu<\omega\le\pi \tag{A6b}$$

$$E\{[v_n^*(\omega)-v_n^*(\nu)]^2\}=N(\omega-\nu)[\pi-(\omega-\nu)],\quad 0\le\nu<\omega\le\pi \tag{A6c}$$

and, the m-variable distribution of the random variables $z_n^*(\omega_1),\ldots, z_n^*(\omega_m)$, $0\le\omega_1<\ldots<\omega_m\le\pi$ is Gaussian, with:

$$P\{z_n^*(0)=0\}=1,\ P\{z_n^*(\pi)=0\}=1 \tag{A7a}$$

$$E\{z_n^*(\omega)\}=0,\quad 0\le\omega\le\pi \tag{A7b}$$

$$E\{[z_n^*(\omega)-z_n^*(\nu)]^2\}=\frac{(\omega-\nu)[\pi-(\omega-\nu)]}{4\pi^2},\quad 0\le\nu<\omega\le\pi \tag{A7c}$$

Equations (A7a) – (A7c) imply:

$$E\{[z_n^*(\omega)]^2\}=\frac{\omega(\pi-\omega)}{4\pi^2} \tag{A6a}$$

$$E\{z_n^*(\nu)z_n^*(\omega)\}=\frac{\nu(\pi-\omega)}{4\pi^2},\quad 0\le\nu<\omega\le\pi \tag{A8b}$$

Equation (a8) indicates that the first term on the right hand side of equation (A2) is in fact $z_n^*(\omega)$.

Let us now consider the distribution of the second term on the right hand side. It can be observed, when $N$ approaches infinite, that:

$$E\{s(\omega)s(\nu)\} = \underset{N\to\infty}{Lim} E\left\{\left(\sum_{\tau=1}^{N-1}\frac{C_\tau}{\pi N^{1/2}}\frac{\sin\tau\omega}{\tau}\right)\left(\sum_{\tau=1}^{N-1}\frac{C_\tau}{\pi N^{1/2}}\frac{\sin\tau\nu}{\tau}\right)\right\}$$
$$= \underset{N\to\infty}{Lim}\frac{1}{\pi^2}\sum_{\tau=1}^{N-1}\frac{\sin\tau\omega}{\tau}\frac{\sin\tau\nu}{\tau} = \frac{\nu(\pi-\omega)}{2\pi^2},\quad 0\le\nu<\omega\le\pi \tag{A9}$$

$$E\left\{[s(\omega)]^2\right\} = \underset{N\to\infty}{Lim} E\left\{\left(\sum_{\tau=1}^{N-1}\frac{C_\tau}{\pi N^{1/2}}\frac{\sin\tau\omega}{\tau}\right)^2\right\} \underset{N\to\infty}{Lim}\frac{1}{\pi^2}\sum_{\tau=1}^{N-1}\left(\frac{\sin\tau\omega}{\tau}\right)^2 = \frac{\omega(\pi-\omega)}{2\pi^2} \tag{A10}$$

Therefore:

$$s(\omega) = \underset{N\to\infty}{Lim}\sum_{\tau=1}^{N-1}\frac{C_\tau}{\pi N^{1/2}}\frac{\sin\tau\omega}{\tau} = 2z_n^*(\omega) \tag{A11}$$

Finally, bringing the results into equation (A2) yields:

$$\xi(\omega) = 3z_n^*(\omega) \tag{A12}$$

The above proves equation (3).

**Table 1.** Institutional features of sectors

| | A&B | D | E | F | G&H | I | J&K | L-Q |
|---|---|---|---|---|---|---|---|---|
| Regulation/policy | H | L | H | L | L | H | L | H |
| Foreign competition | H | H | L | L | H | L | H | L |
| Demand | H | H | H | H | H | H | H | L |
| Supply | L | H | L | L | L | H | H | H |

H - subject to the influence of the specific factor to a high degree;
L - subject to the influence of the specific factor to a low degree.

**Table 2.** Summary statistics

| | A&B | D | E | F | G&H | I | J&K | L-Q | J-Q | GDP |
|---|---|---|---|---|---|---|---|---|---|---|
| Mean | 0.3810 | 0.3298 | 0.7864 | 0.4682 | 0.5732 | 0.7903 | 0.8945 | 0.4673 | 0.6188 | 0.6001 |
| Std | 2.3046 | 1.7676 | 4.2544 | 2.7480 | 1.4162 | 1.5019 | 0.9018 | 0.3678 | 0.7042 | 1.0121 |
| Acc | 0.0018 | -0.0110 | 0.0012 | -0.0110 | 0.0004 | 0.0051 | -0.0026 | 0.0020 | -0.0010 | 0.0013 |

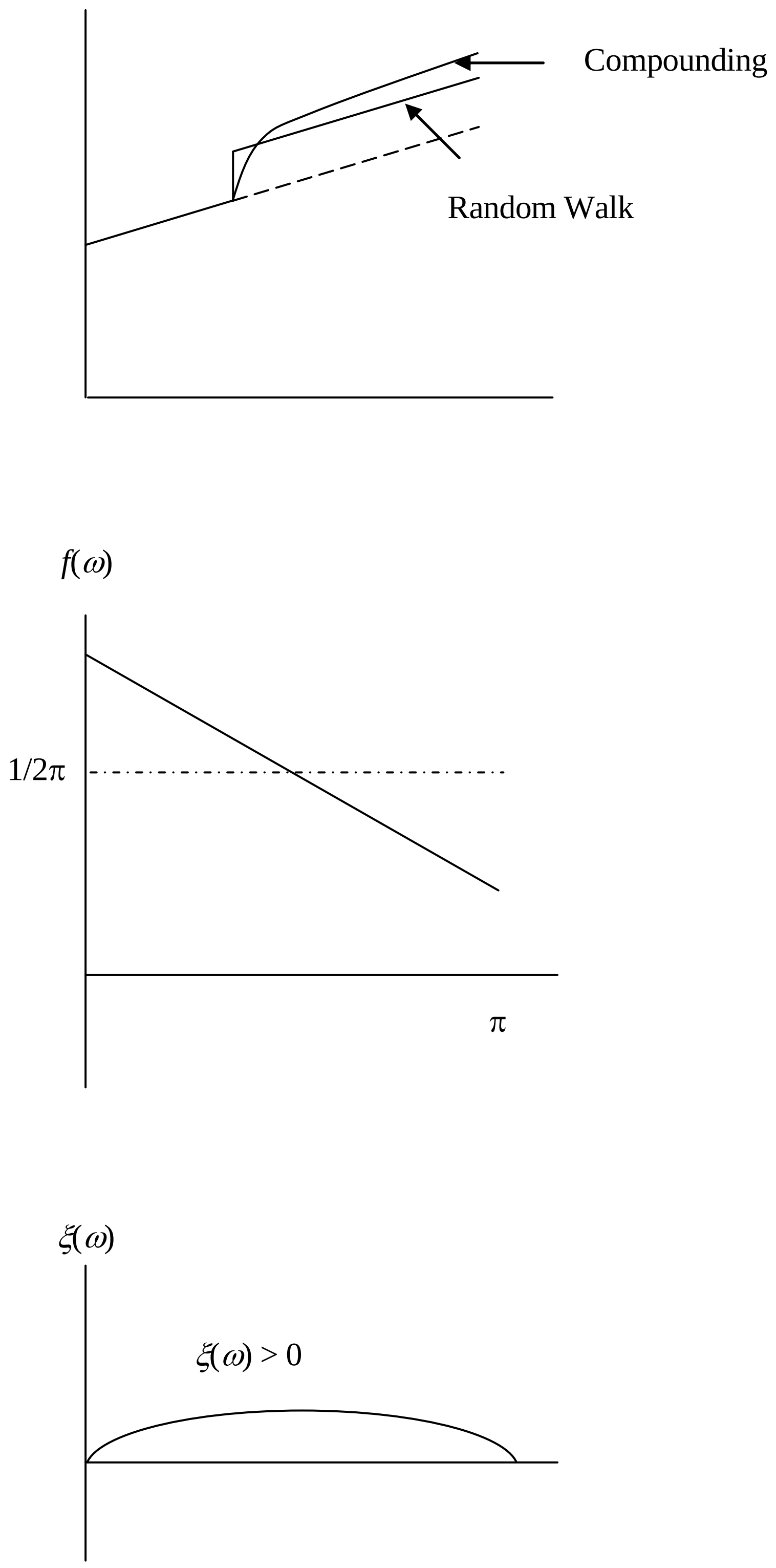

**Figure 1. Lower frequencies dominate (compounding effect)**

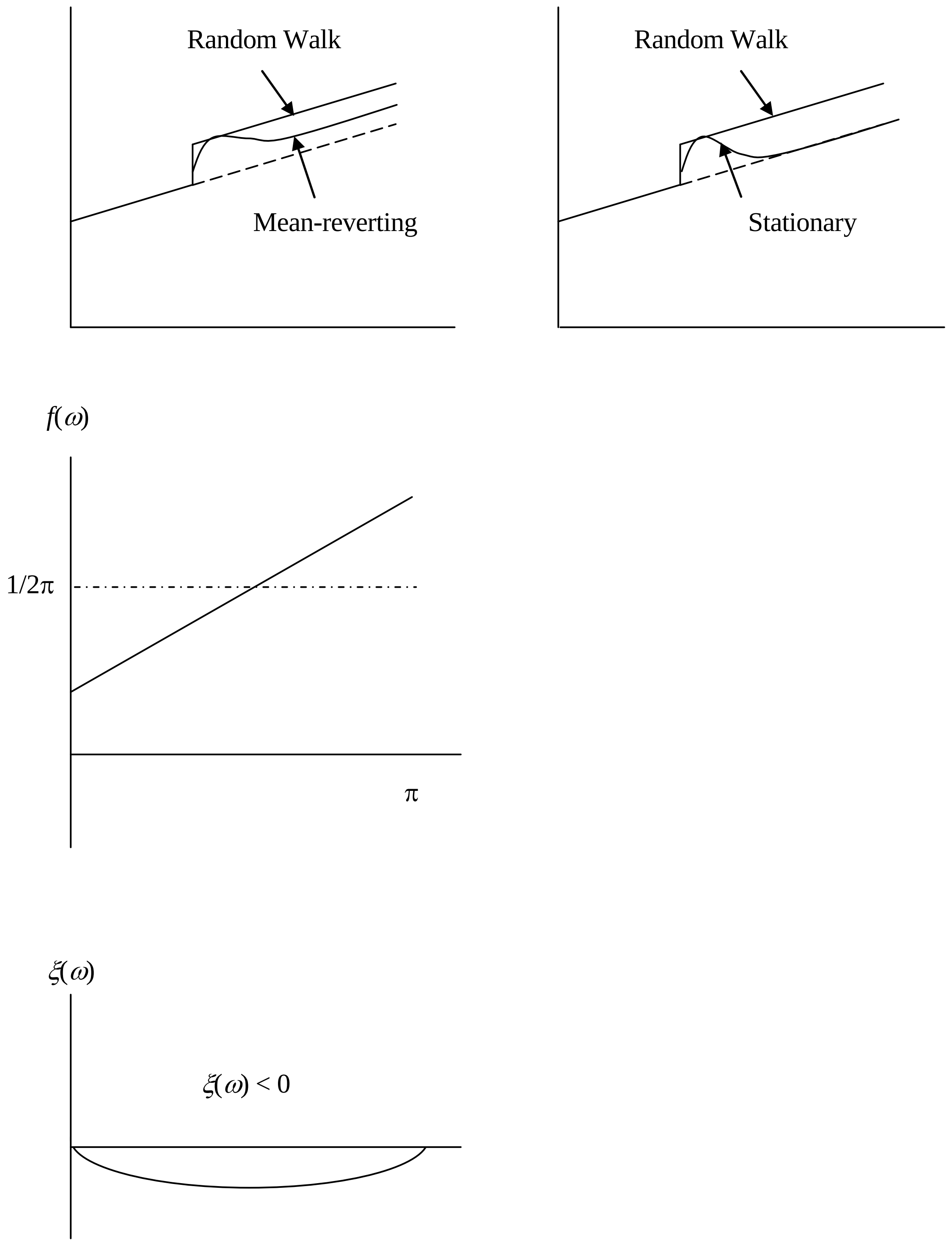

**Figure 2. Higher frequencies dominate (mean-reverting tendency)**

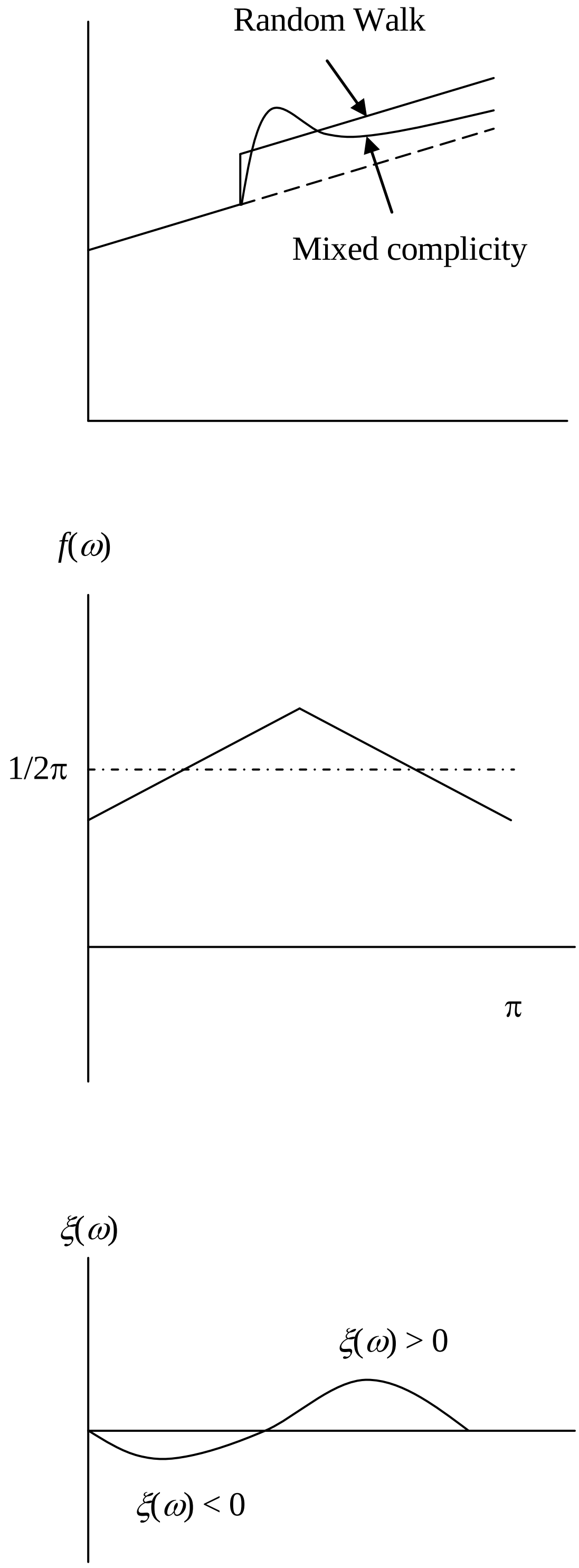

**Figure 3(a). Mixed complicity**

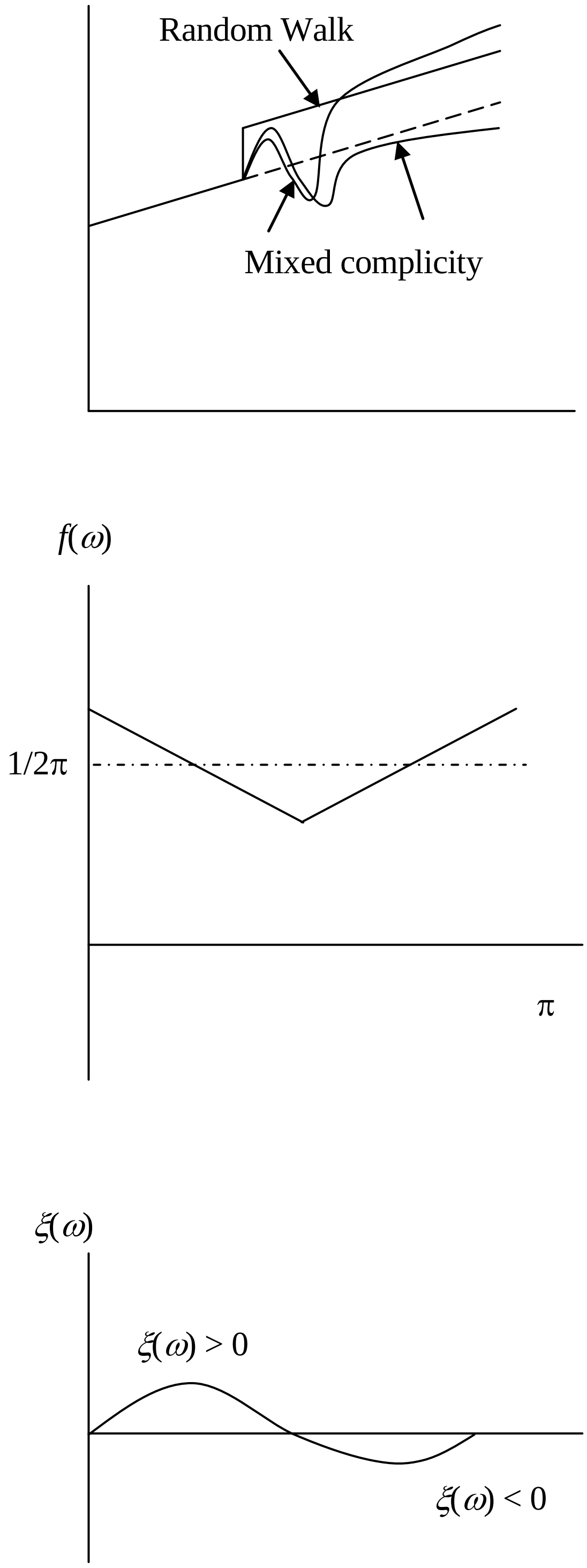

**Figure 3(b). Mixed complicity**

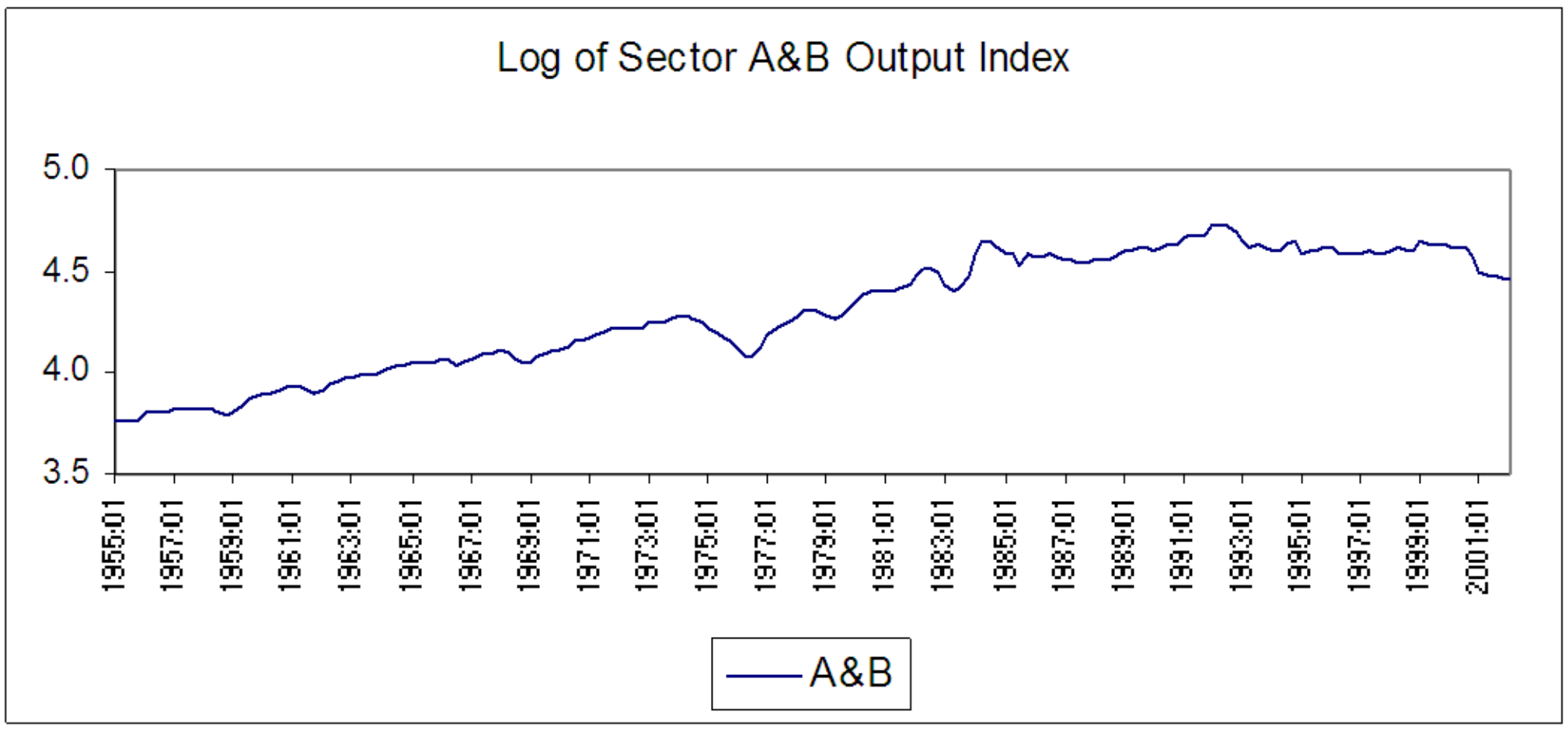

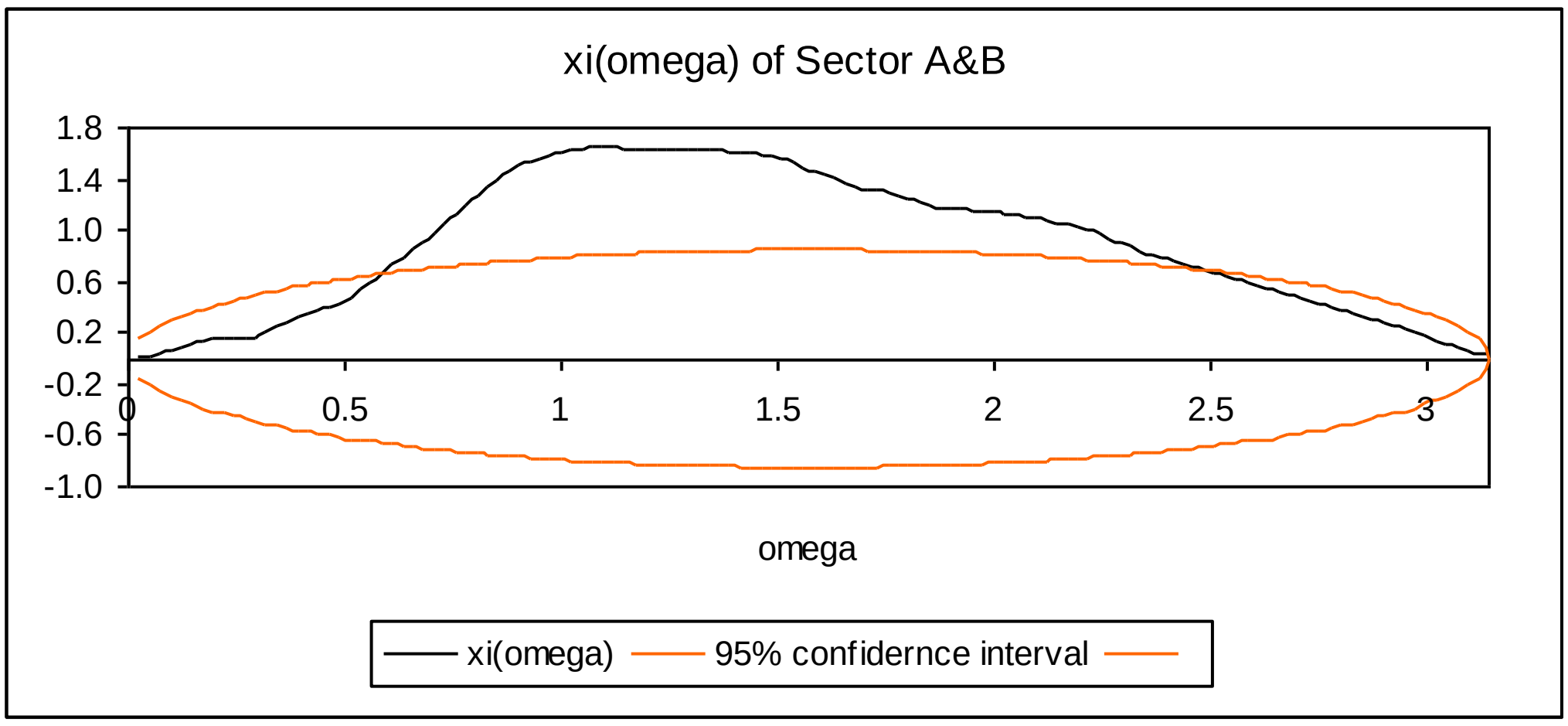

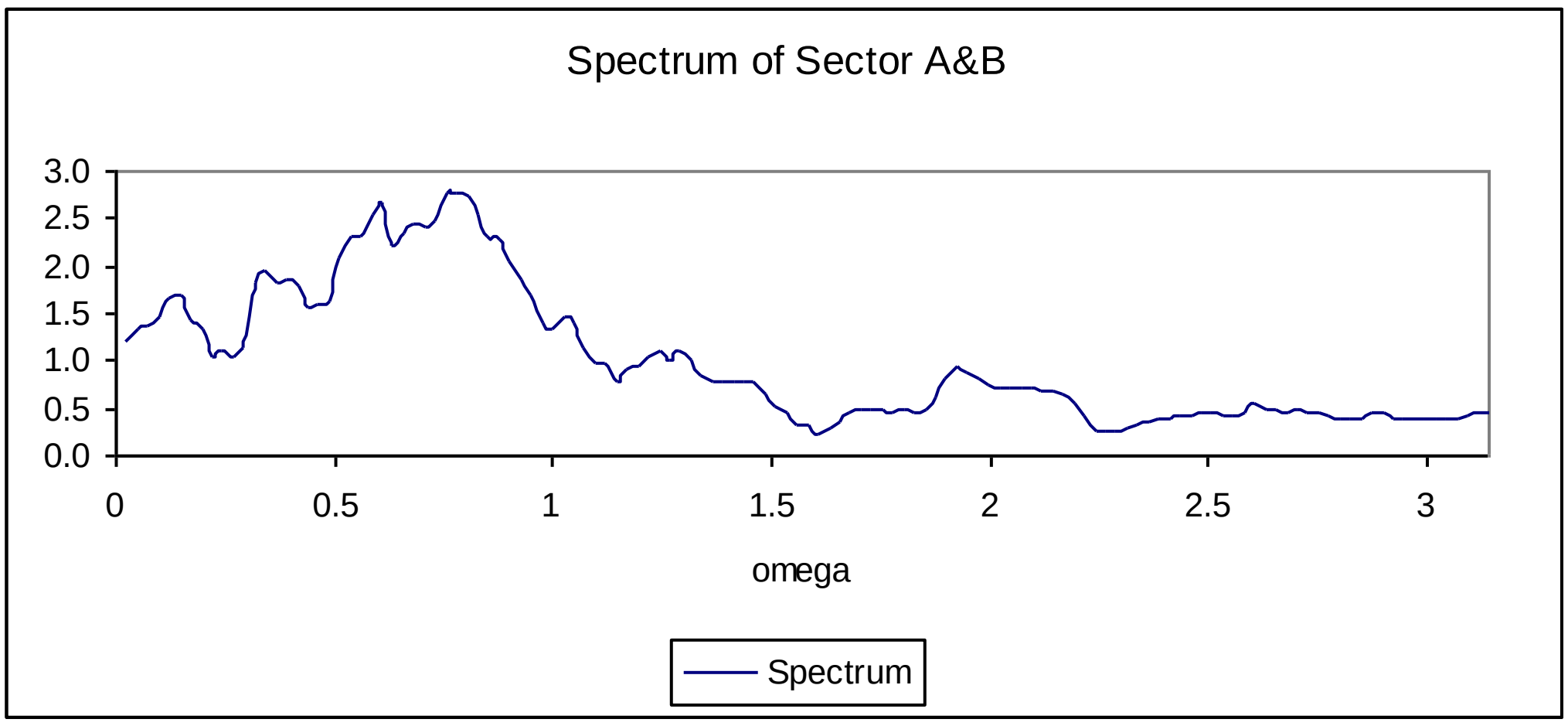

**Figure 4. Business Cycle Patterns: Sector A&B**

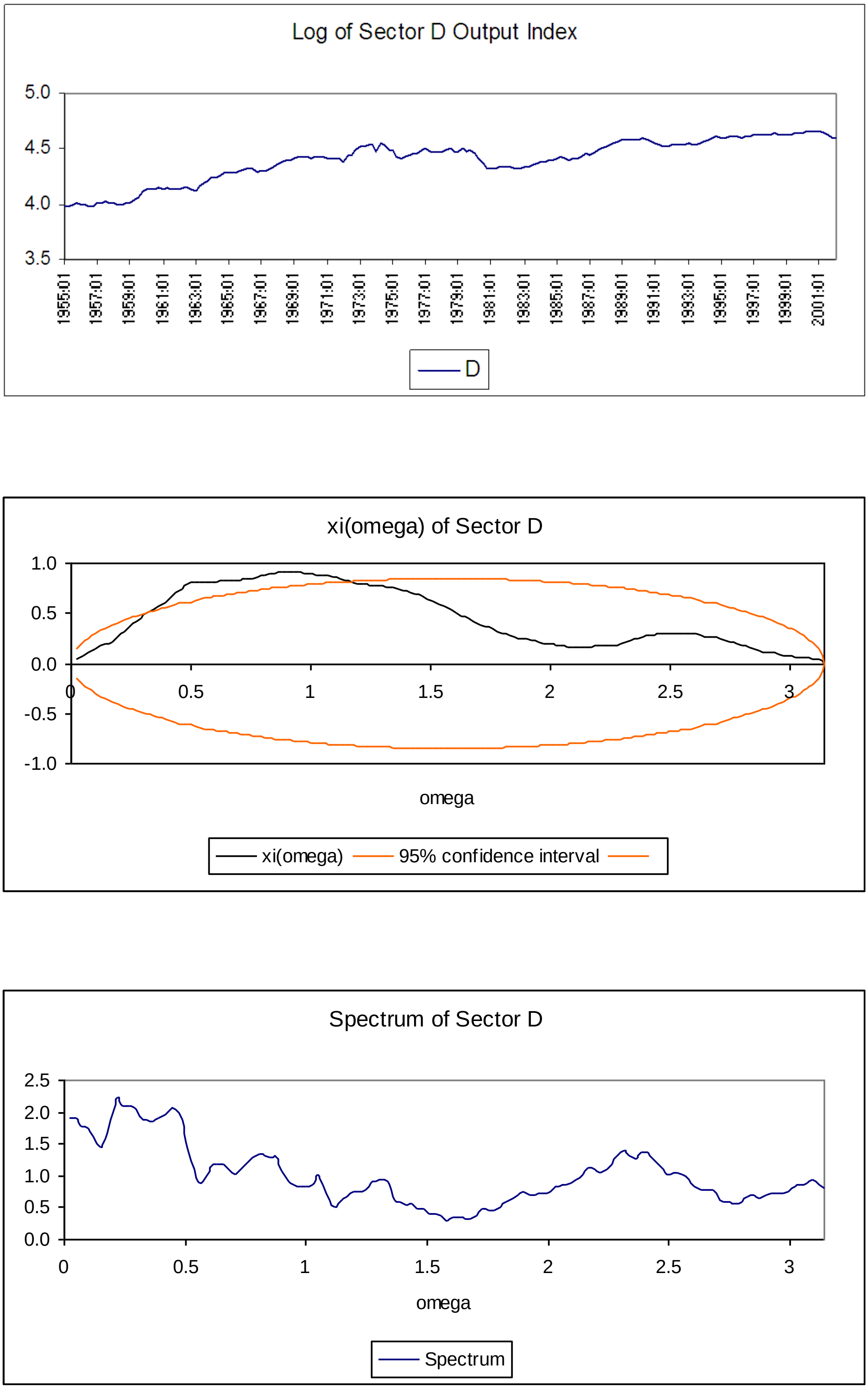

**Figure 5. Business Cycle Patterns: Sector D**

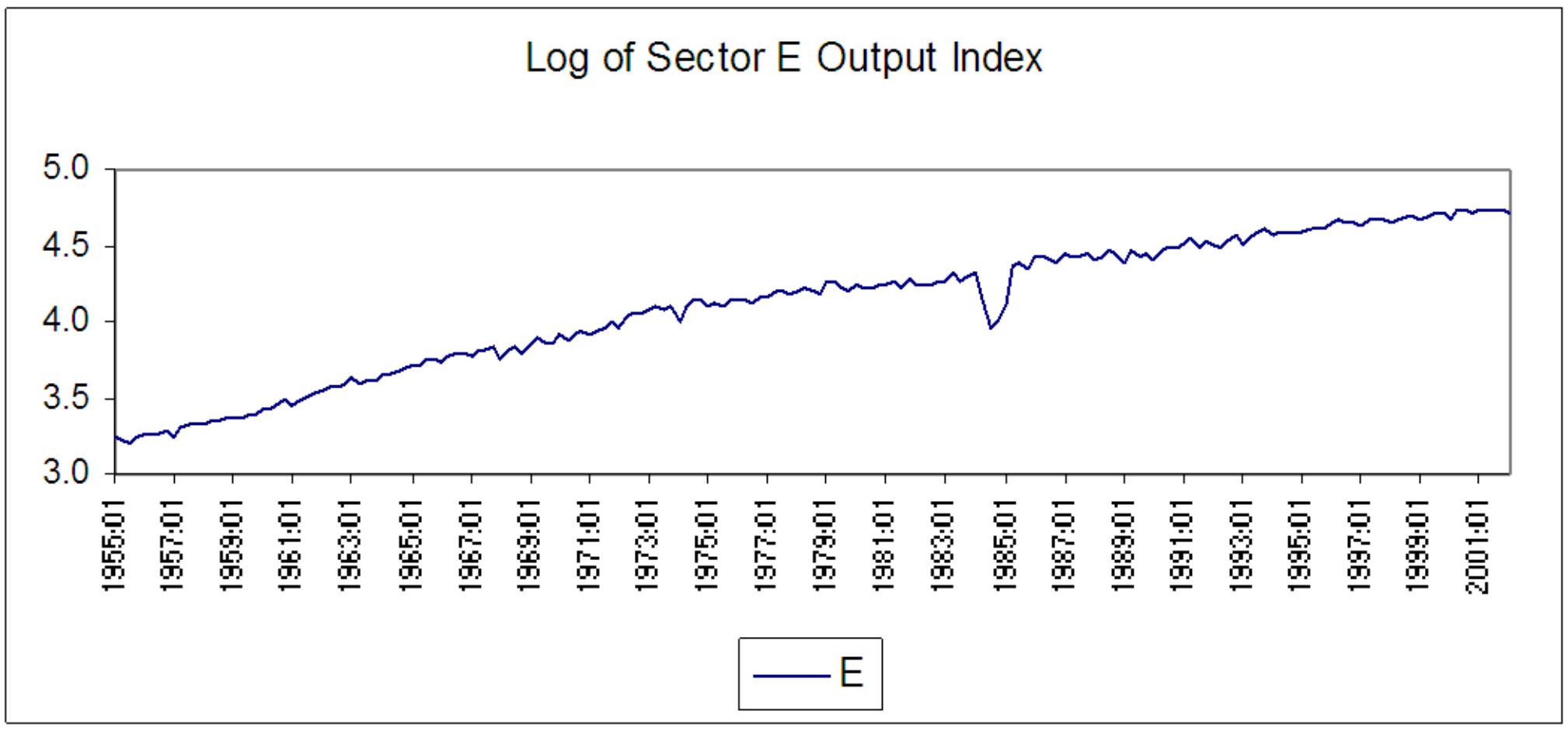

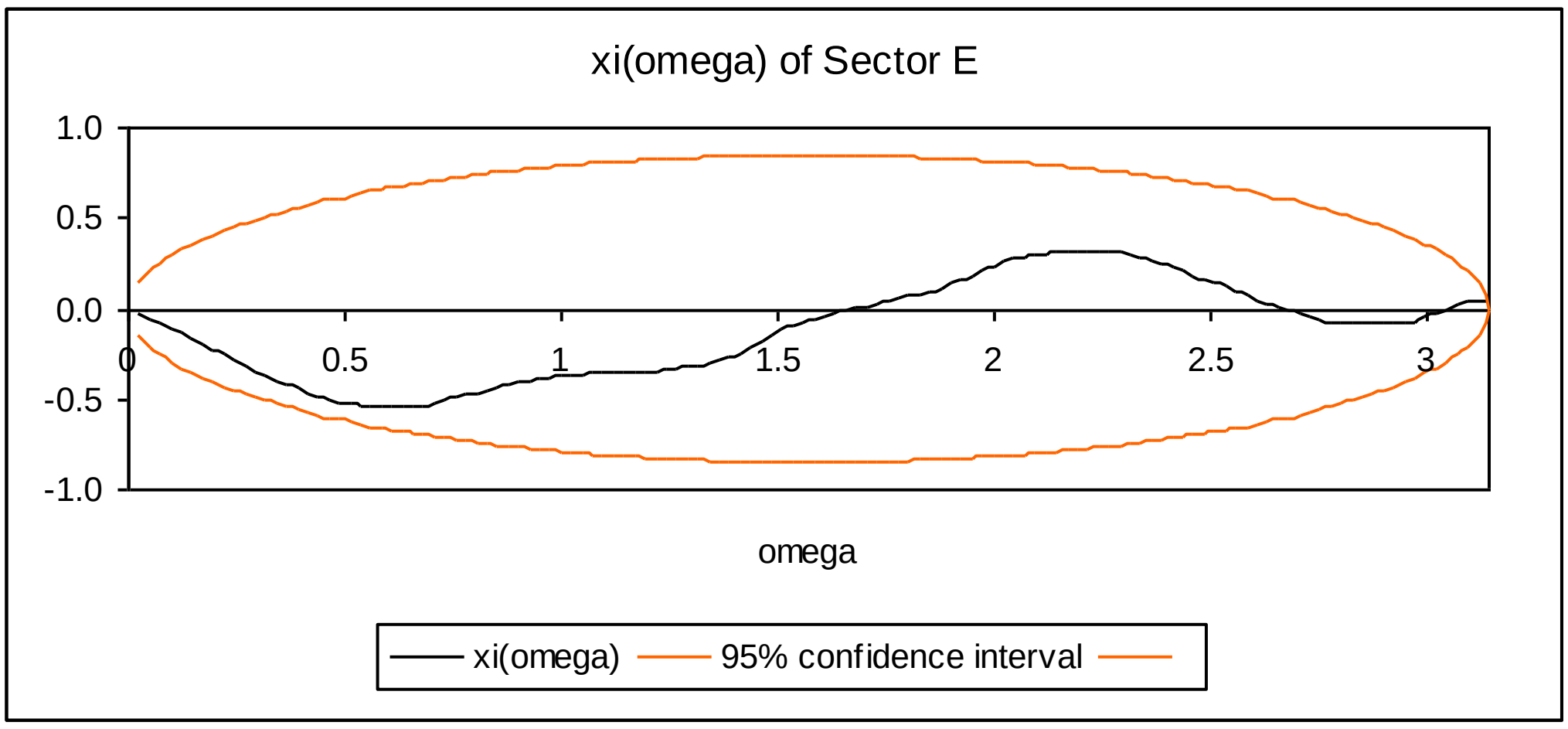

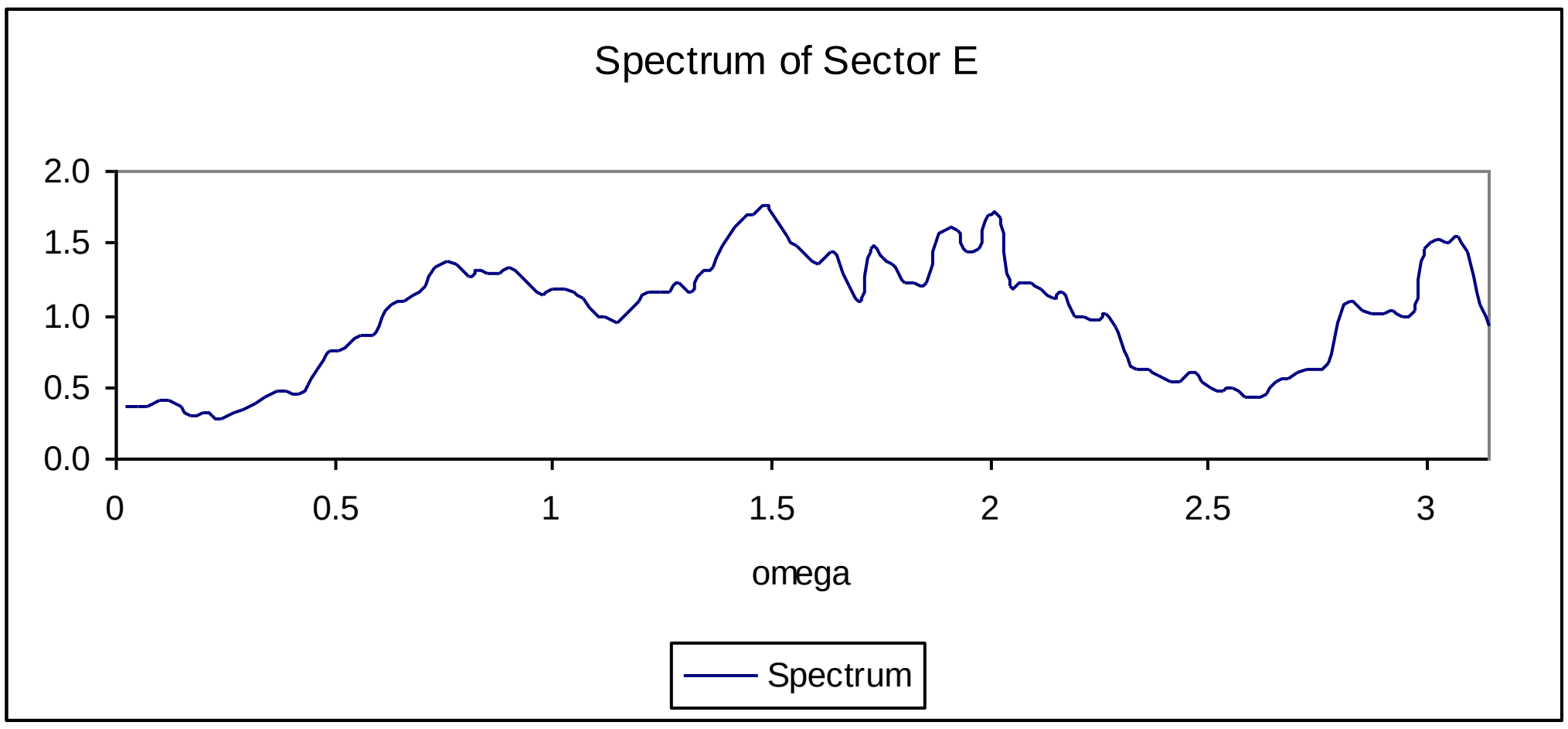

**Figure 6. Business Cycle Patterns: Sector E**

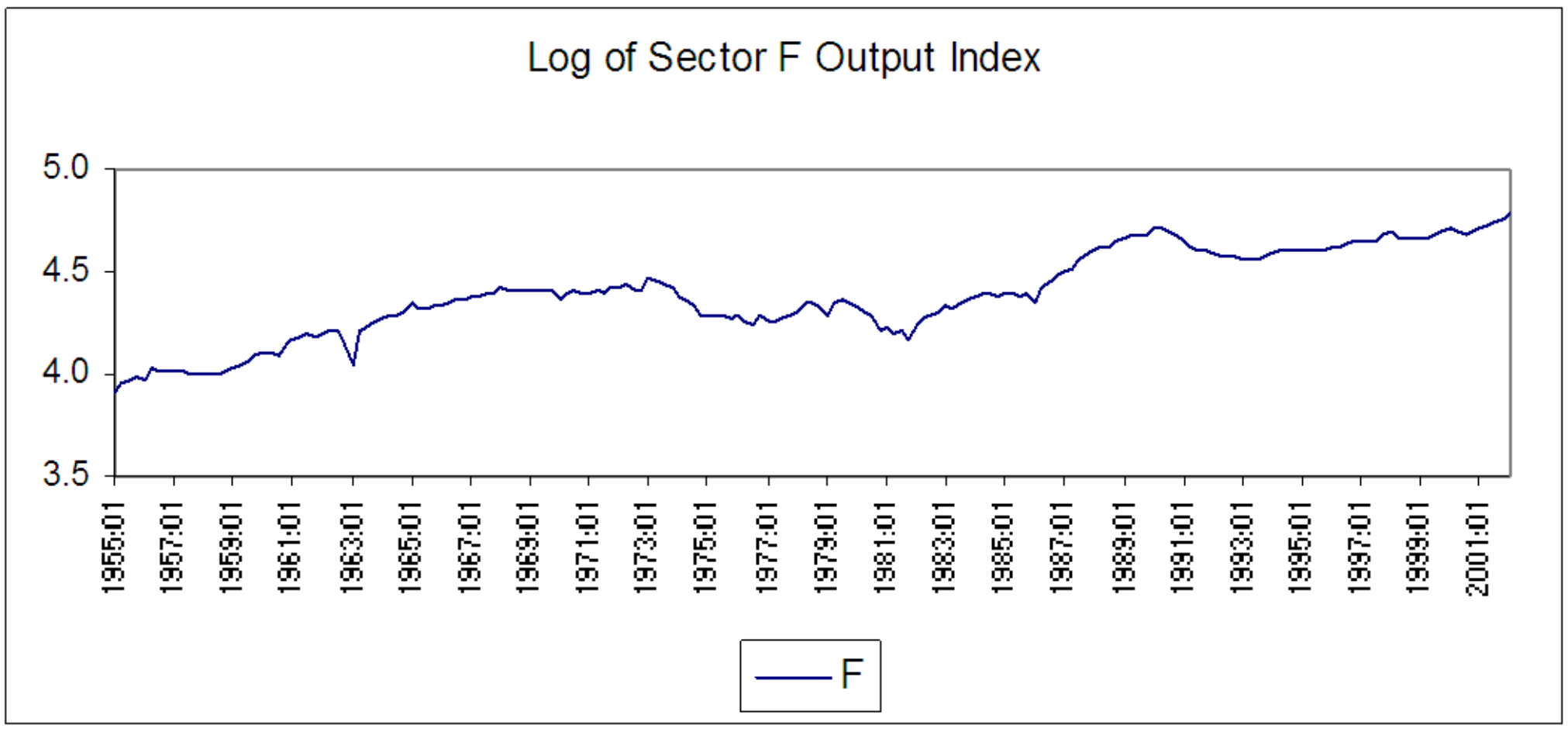

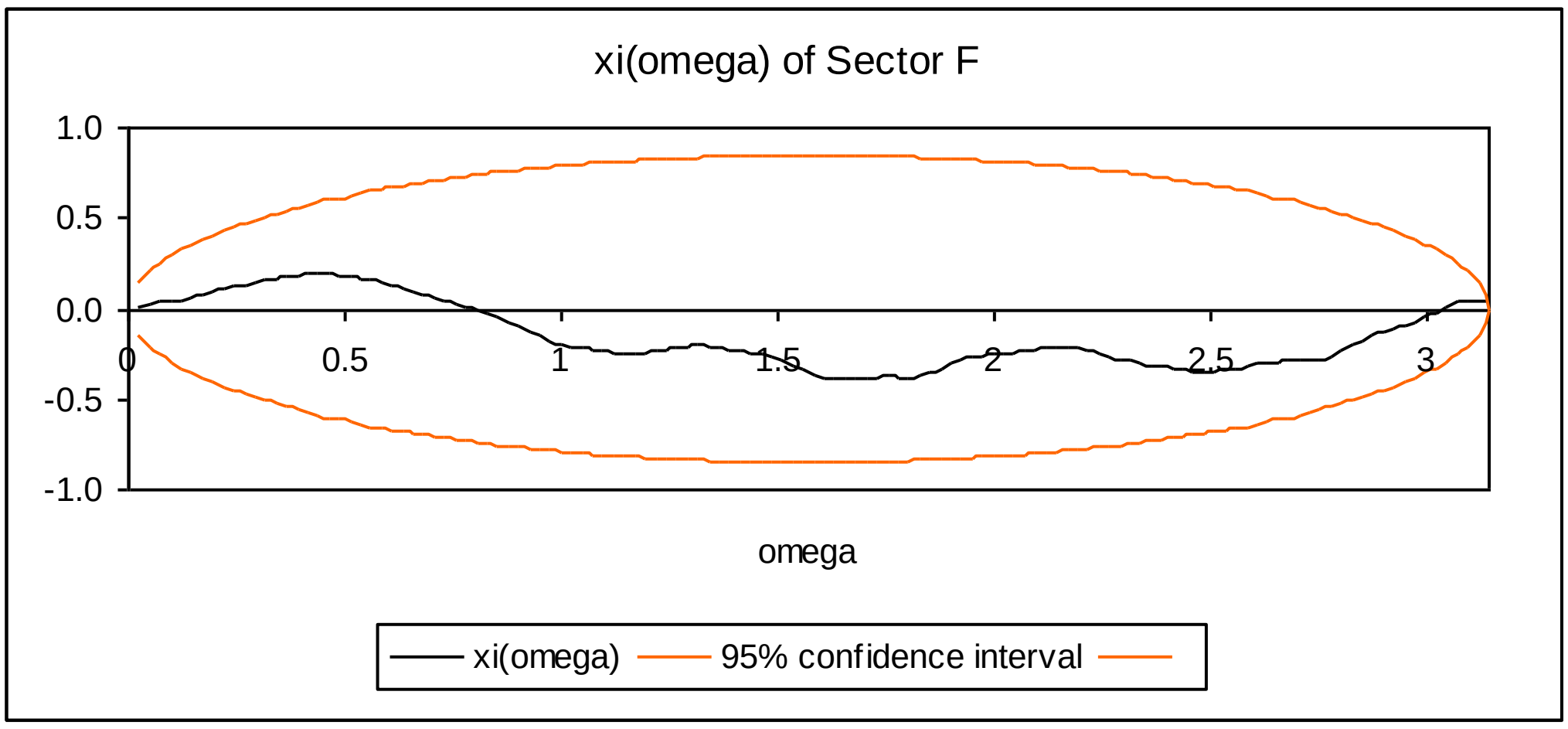

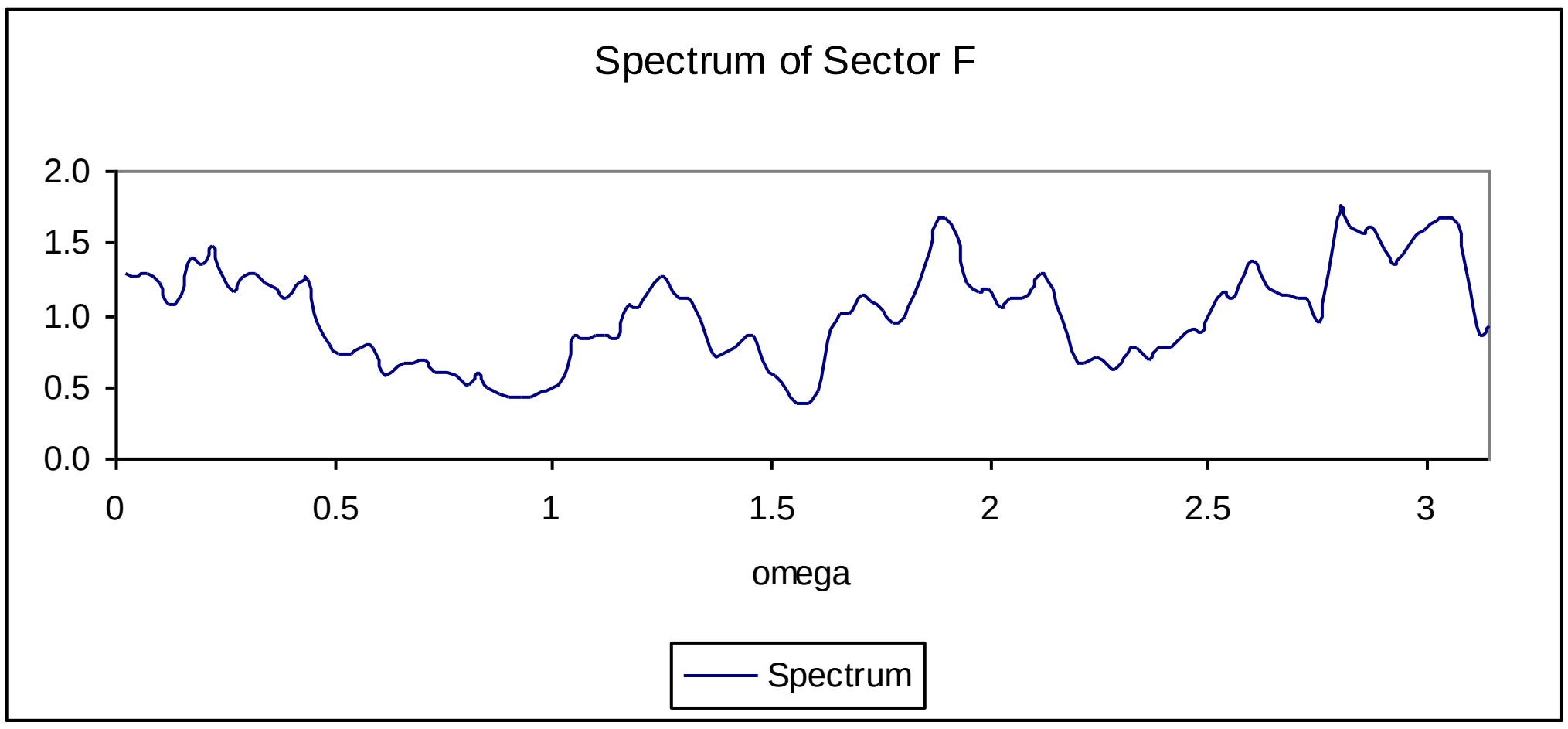

**Figure 7. Business Cycle Patterns: Sector F**

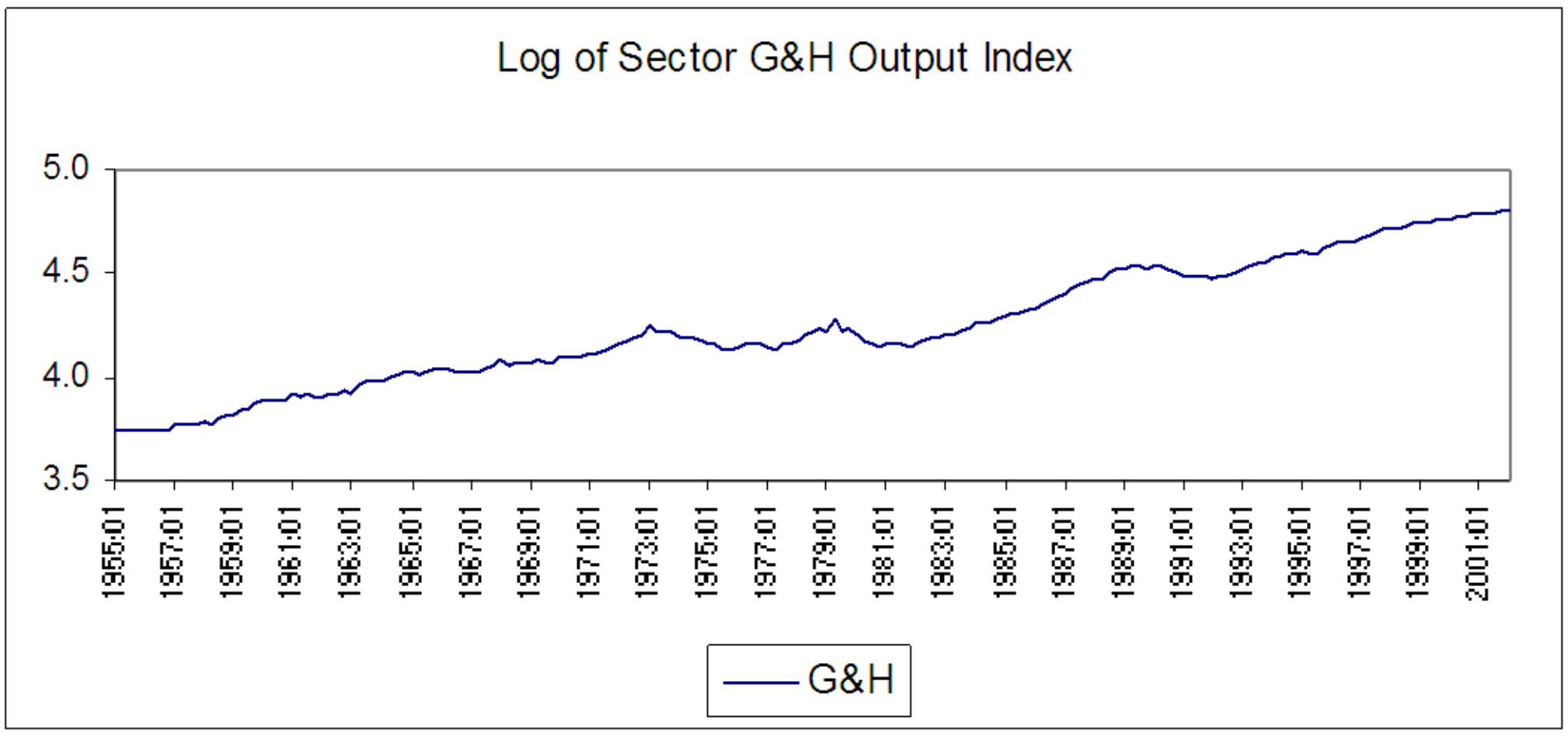

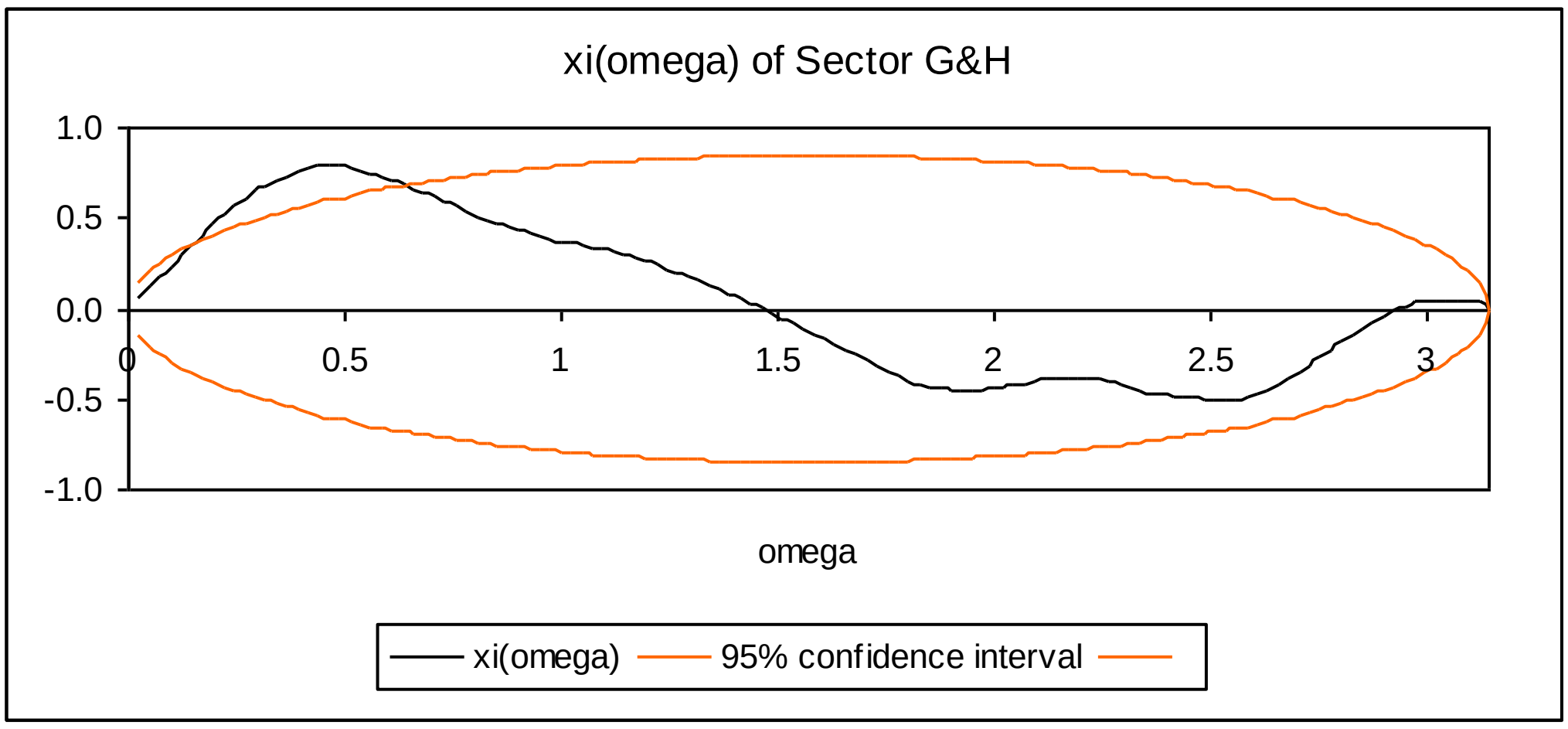

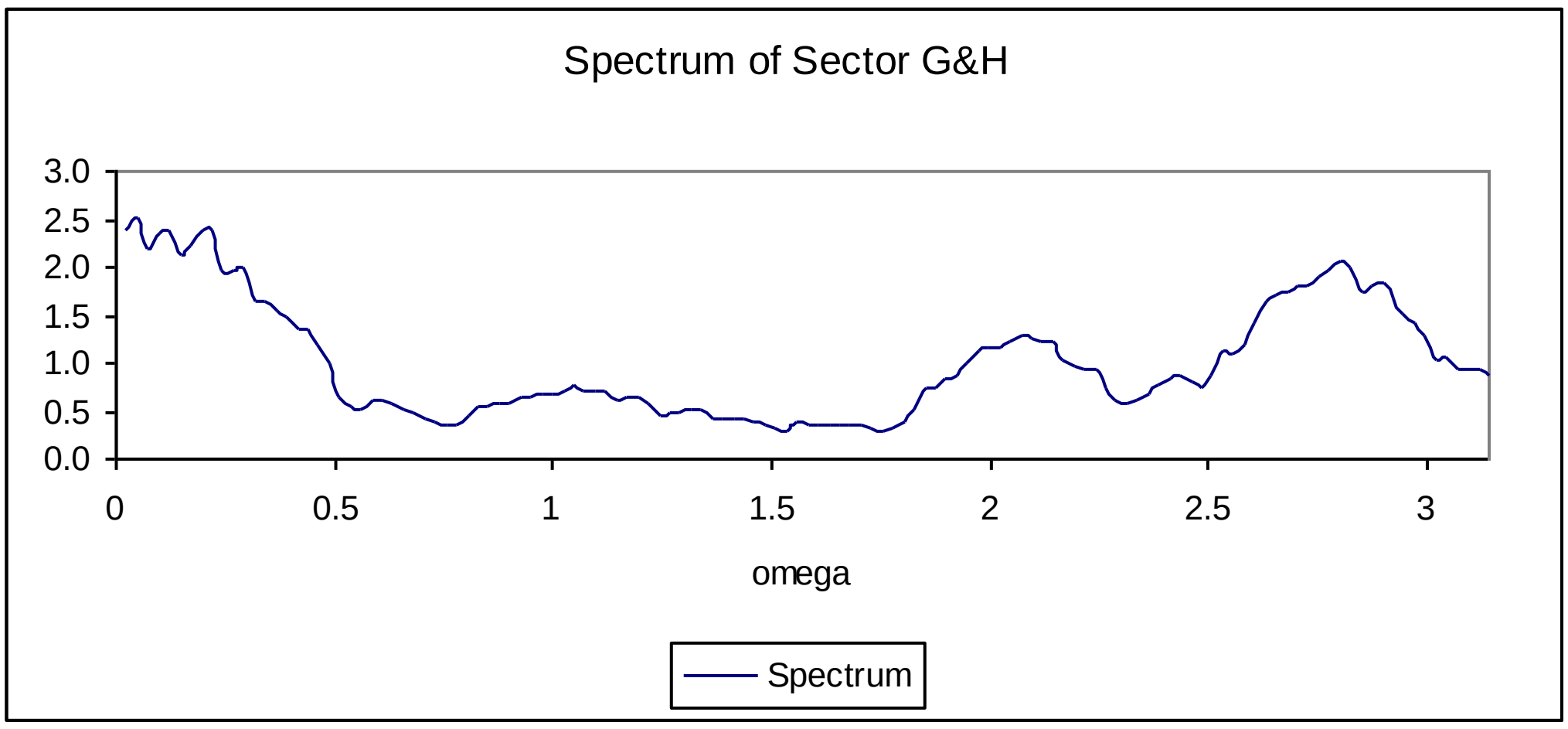

**Figure 8. Business Cycle Patterns: Sector G&H**

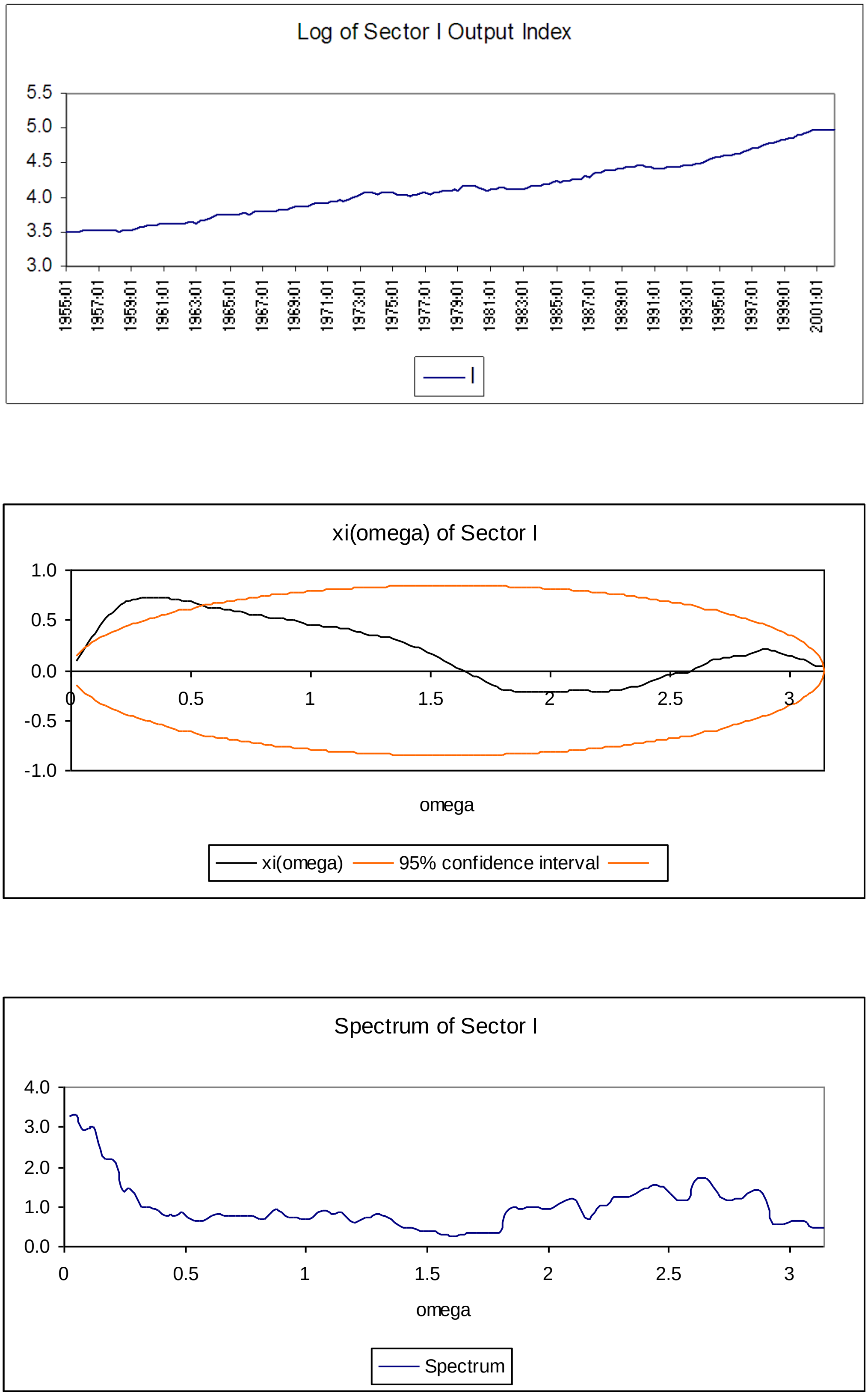

**Figure 9. Business Cycle Patterns: Sector I**

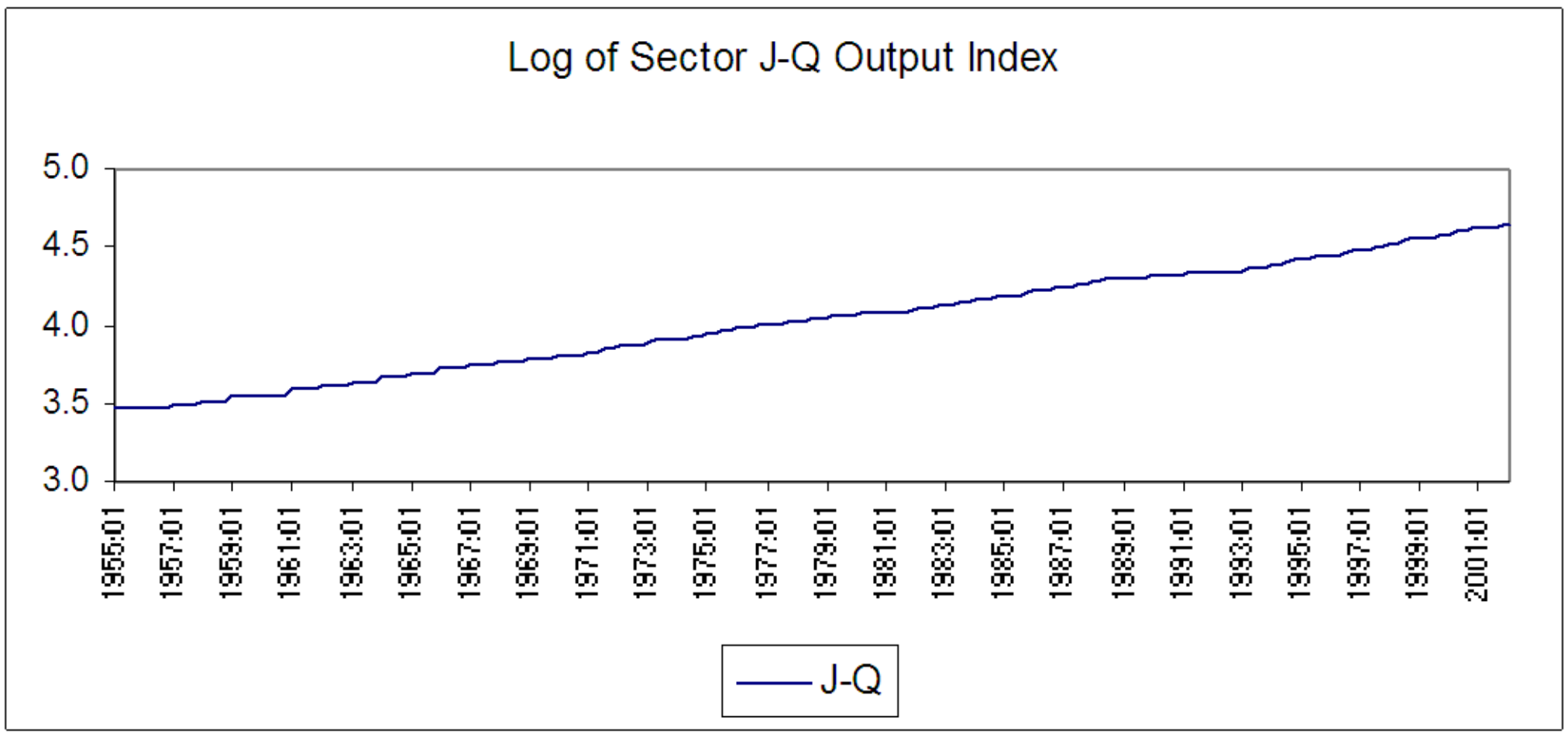

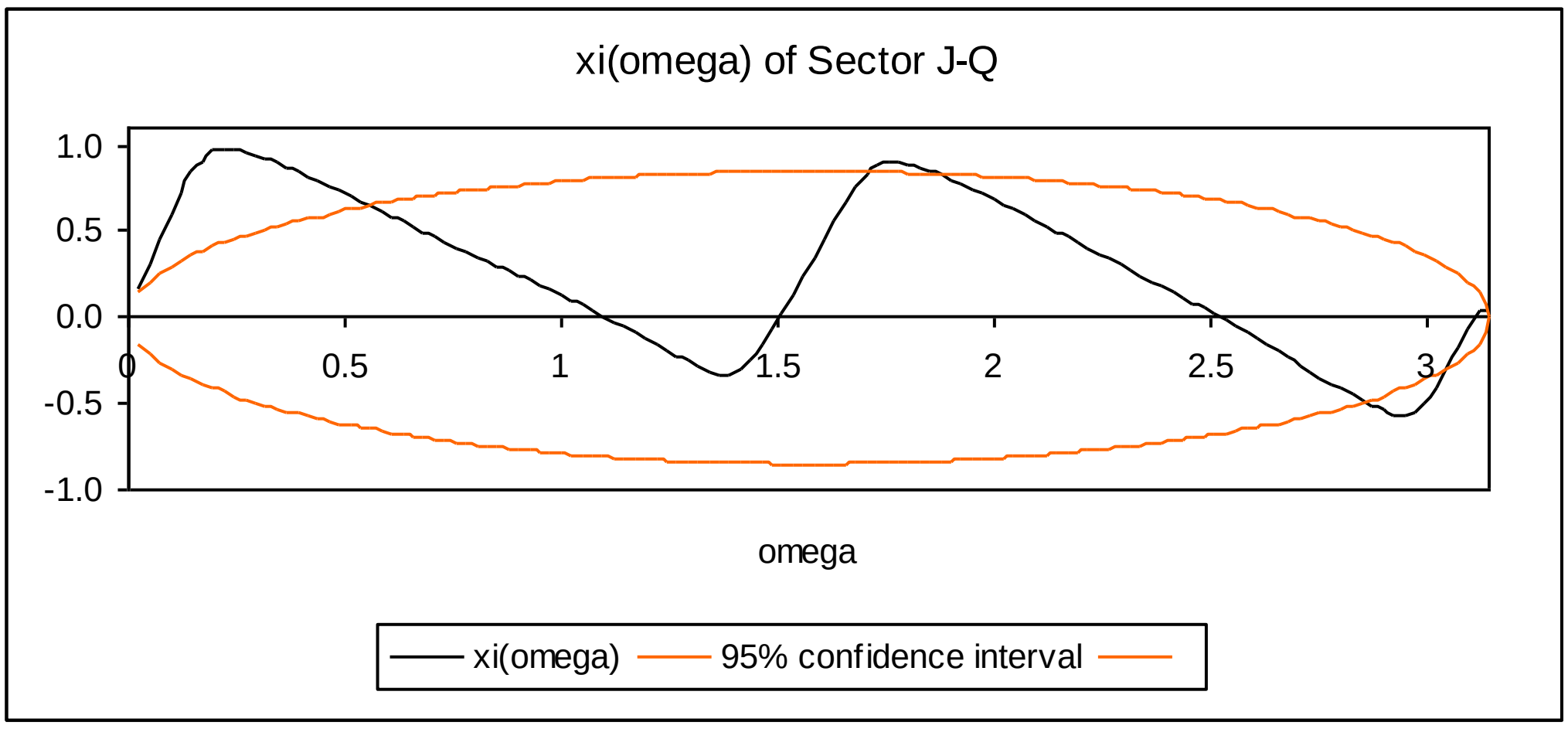

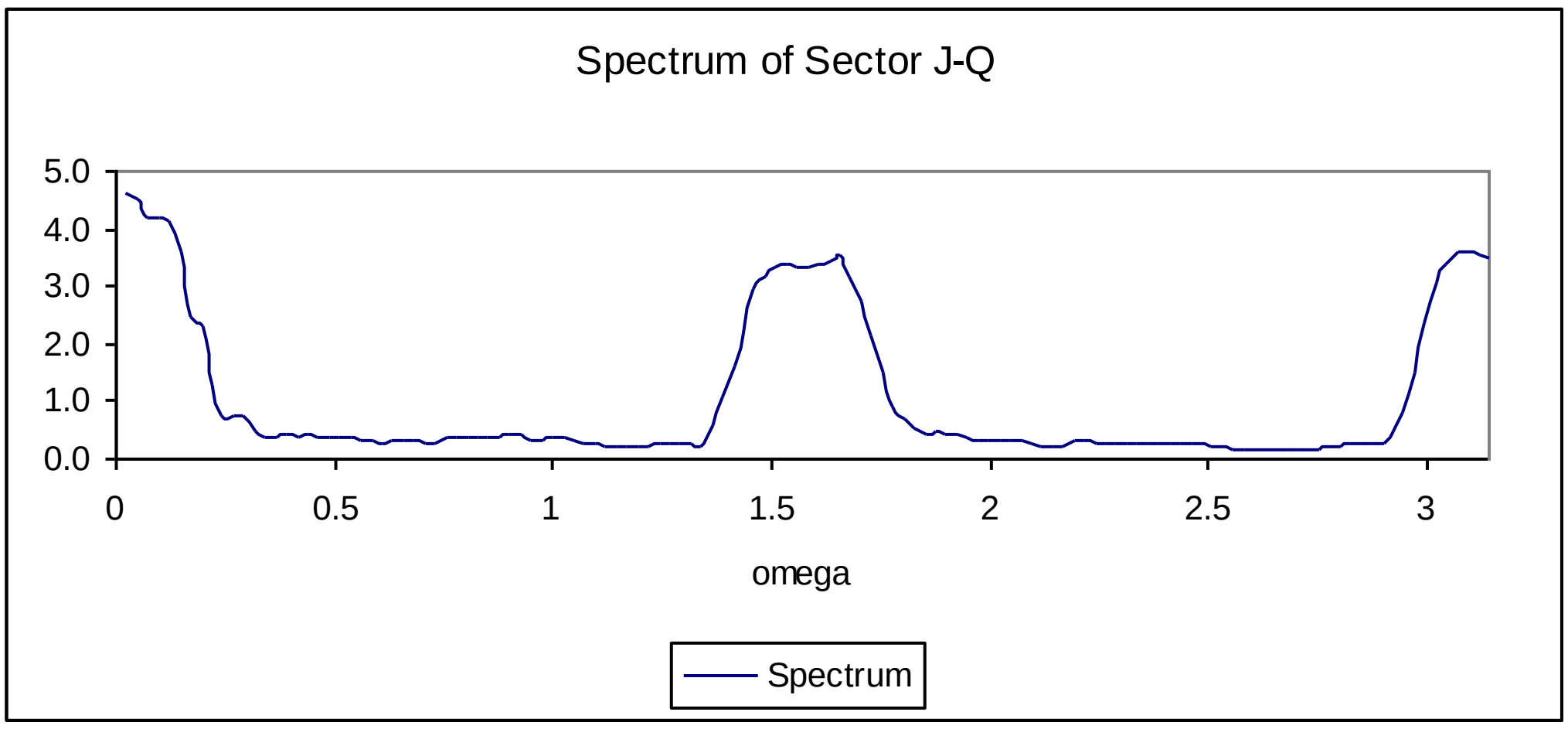

**Figure 10(a). Business Cycle Patterns: Sector J-Q (1955-2002)**

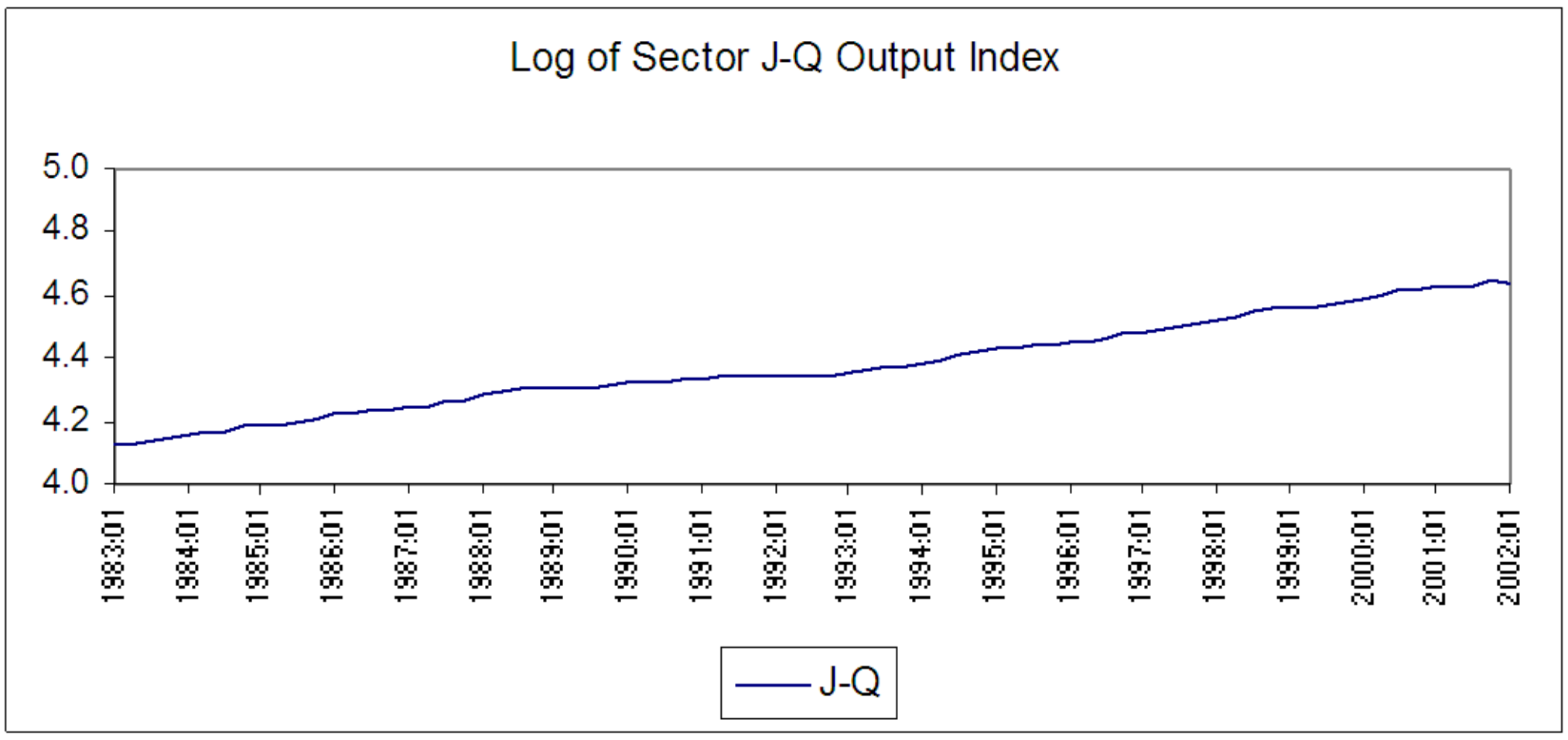

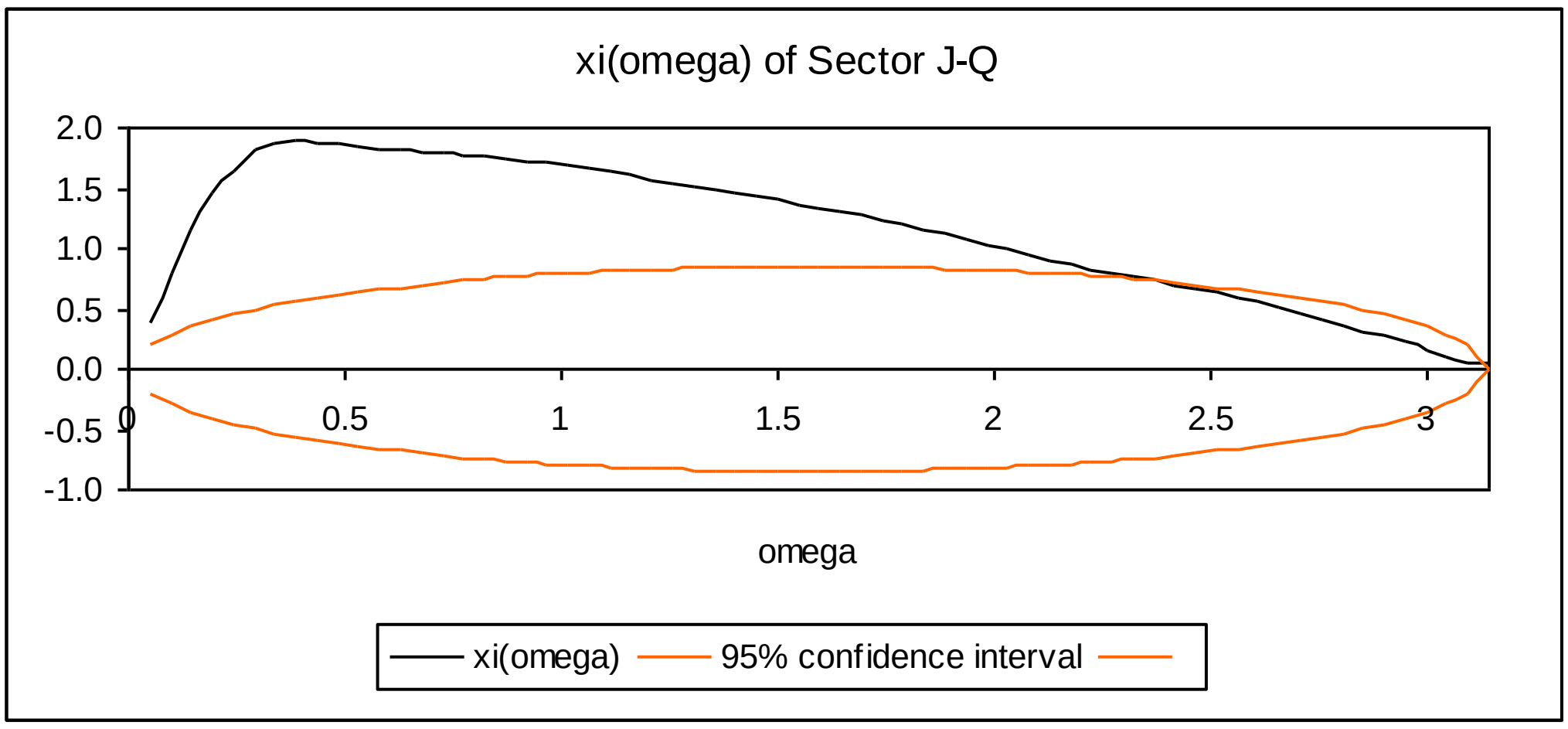

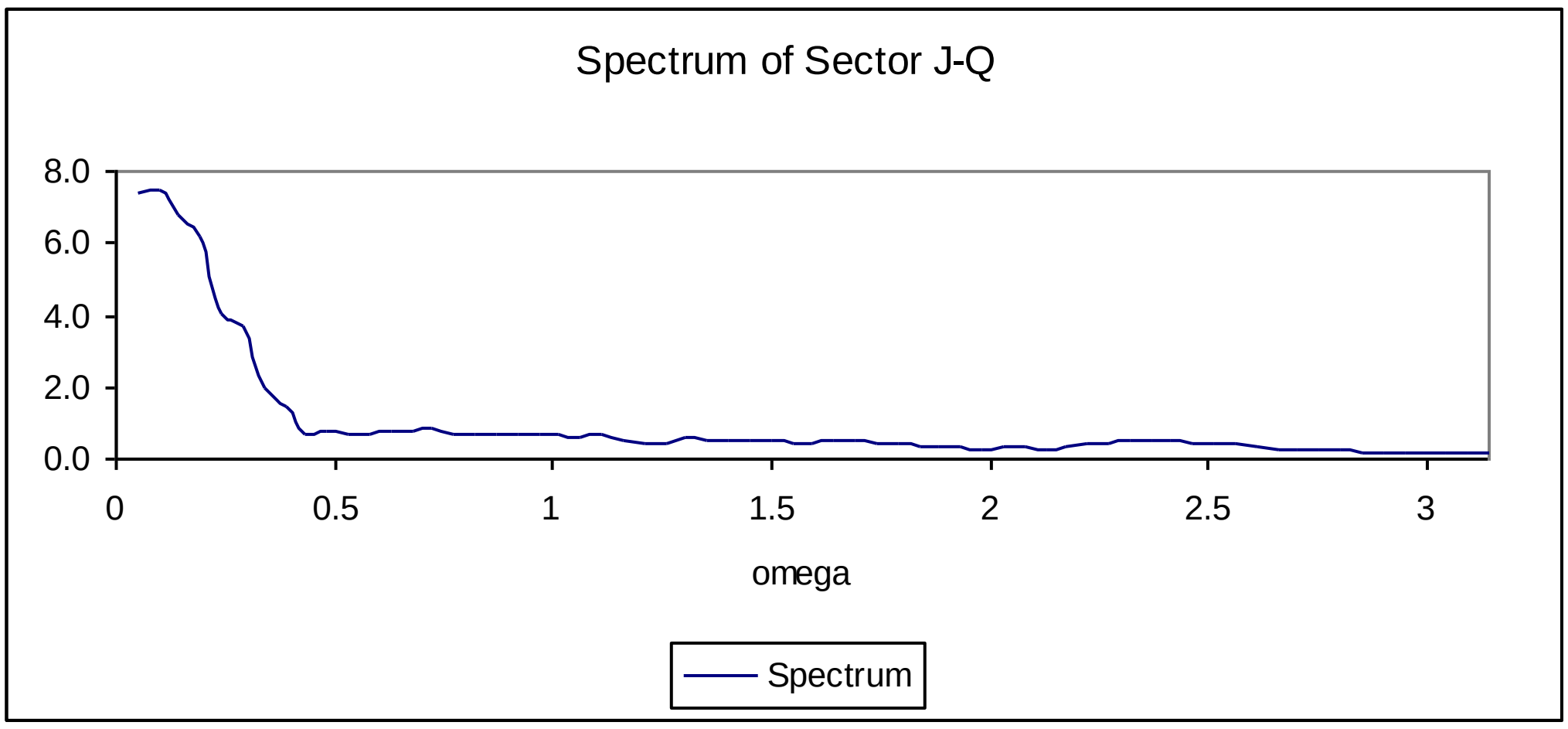

**Figure 10(b). Business Cycle Patterns: Sector J-Q (1983-2002)**

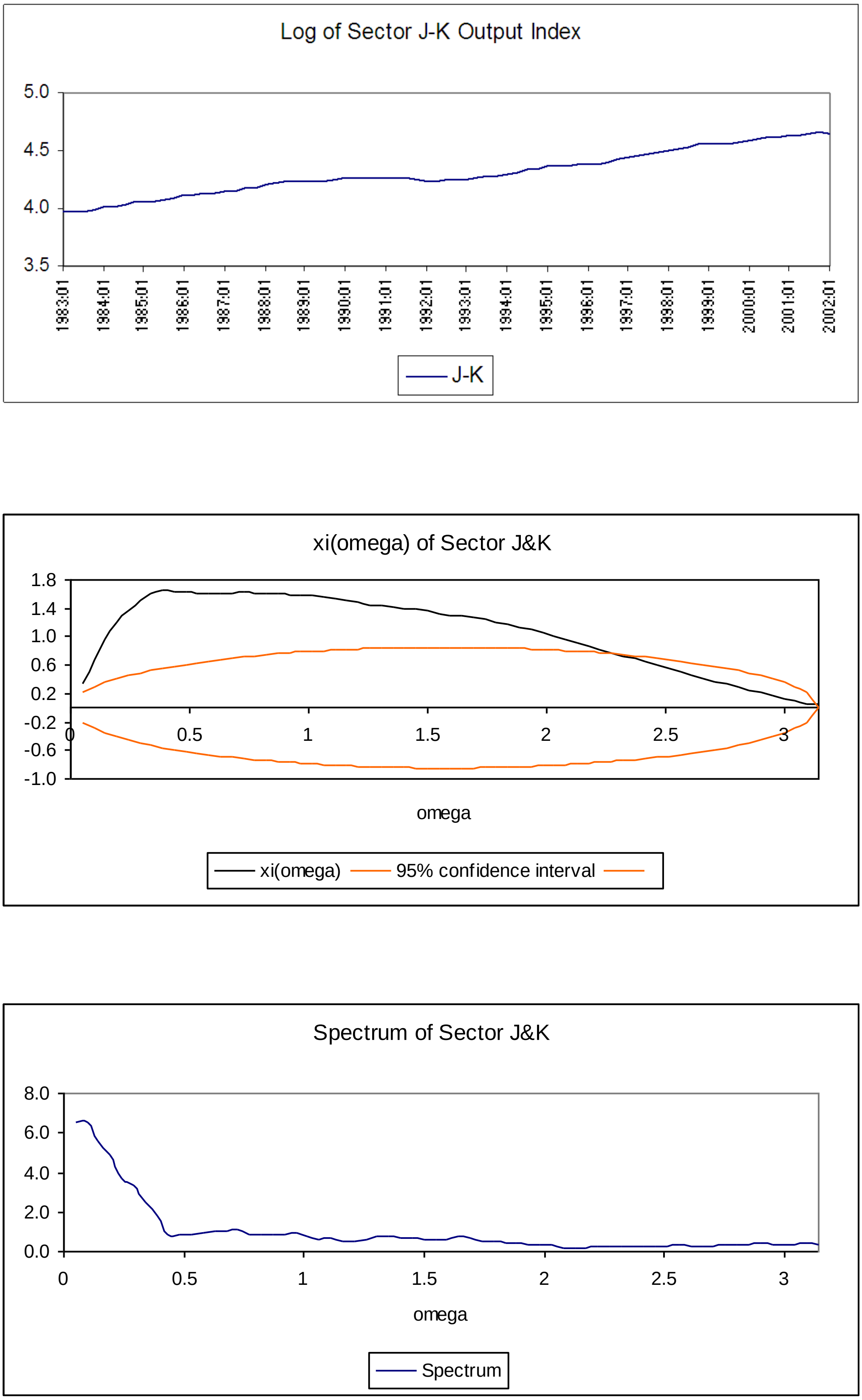

**Figure 11. Business Cycle Patterns: Sector J&K**

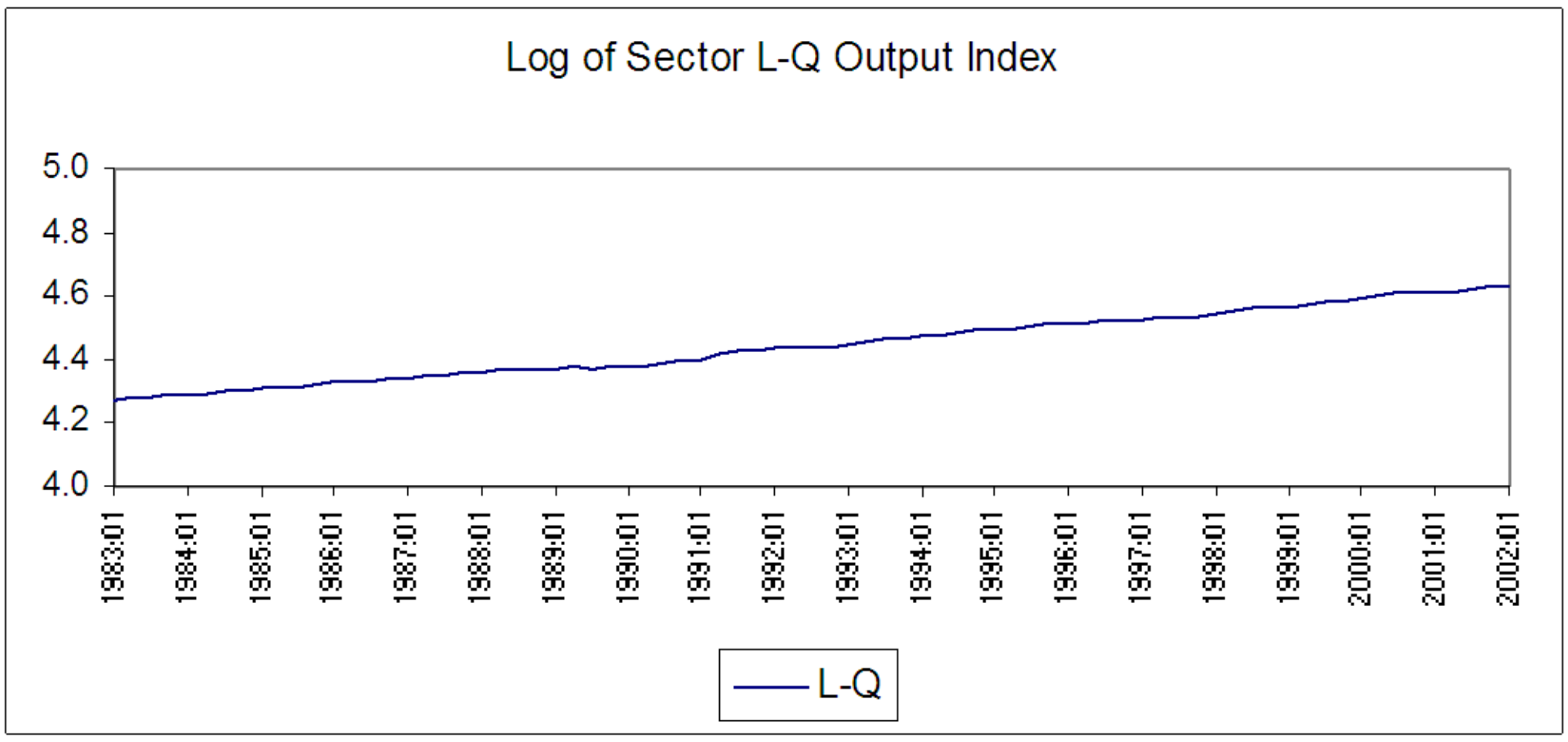

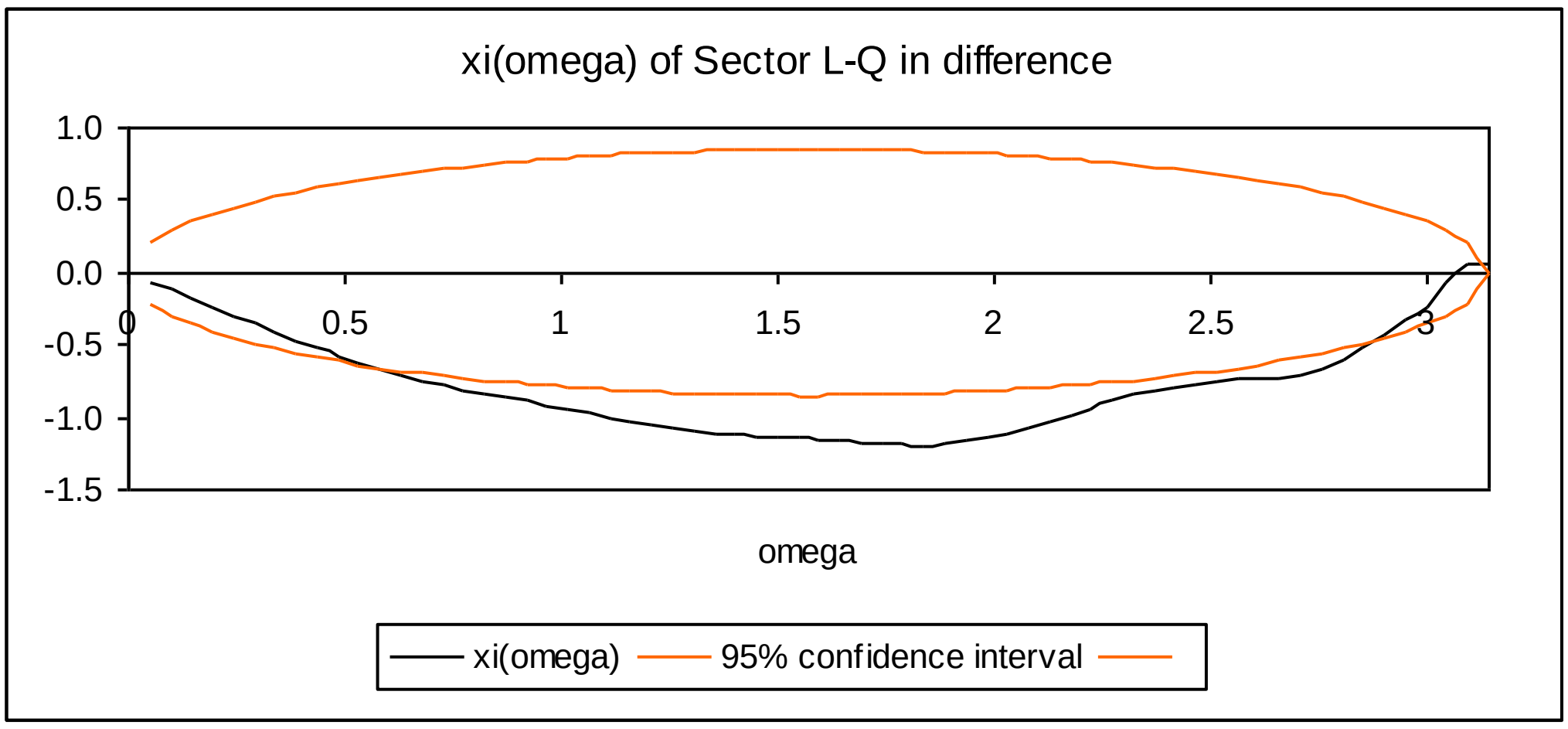

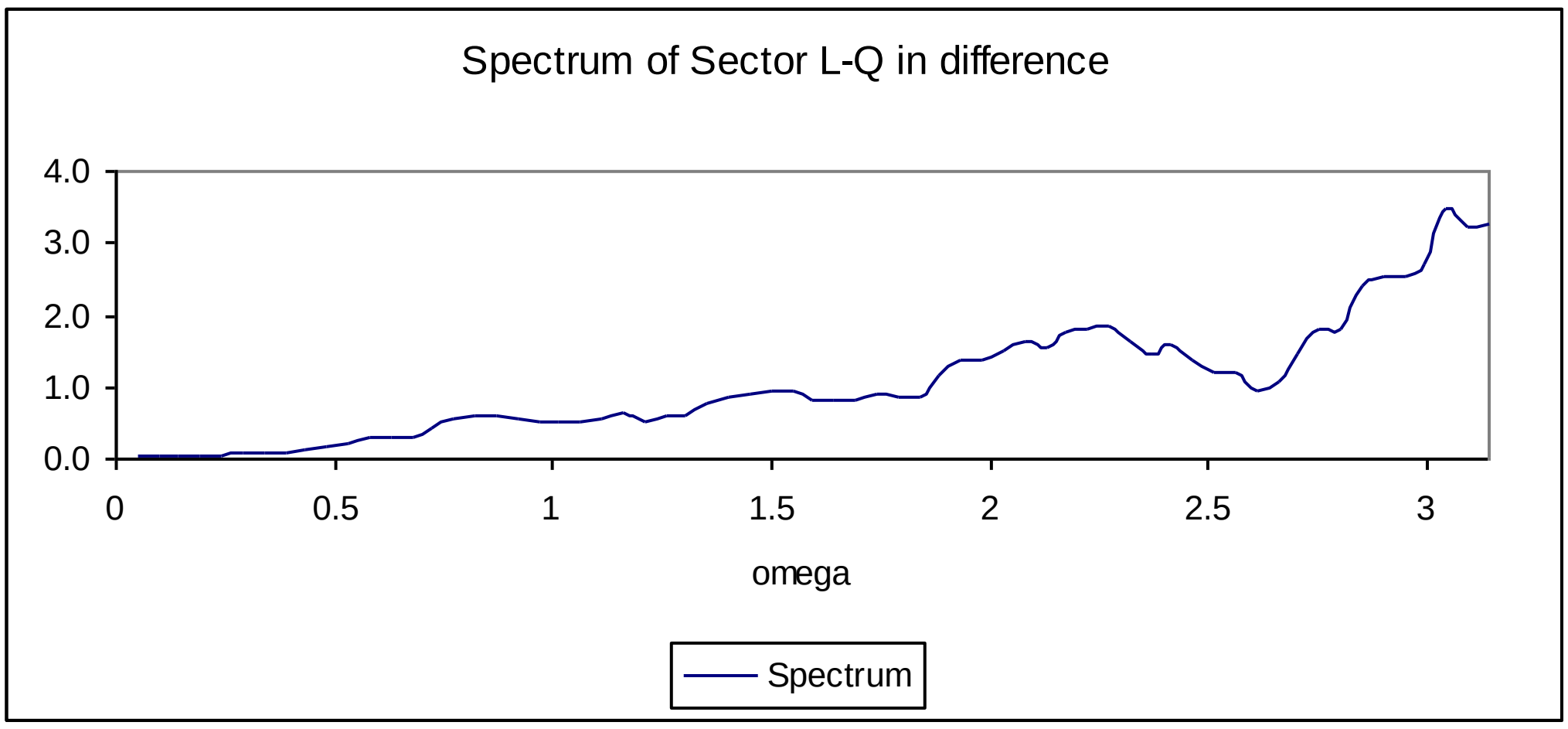

**Figure 12. Business Cycle Patterns: Sector L-Q**

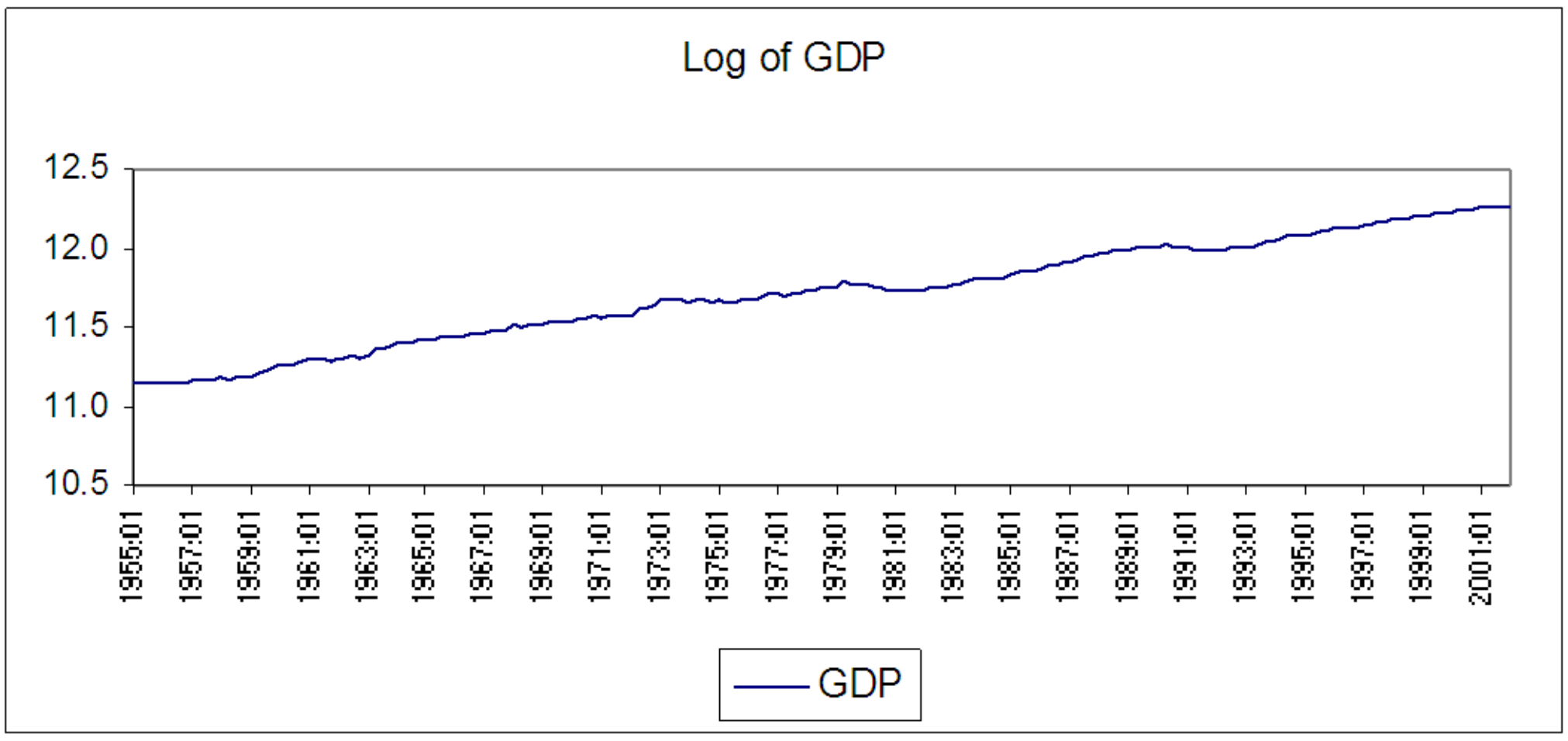

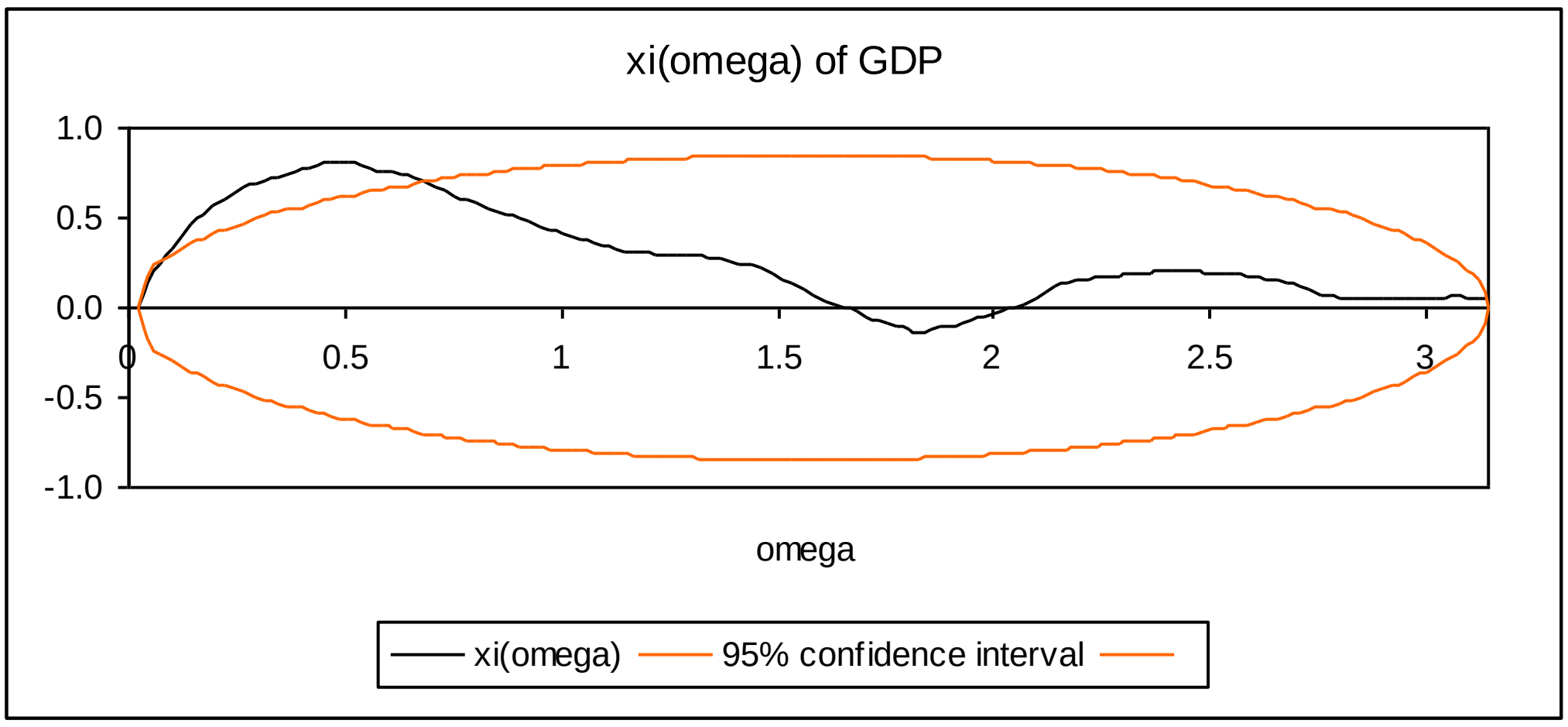

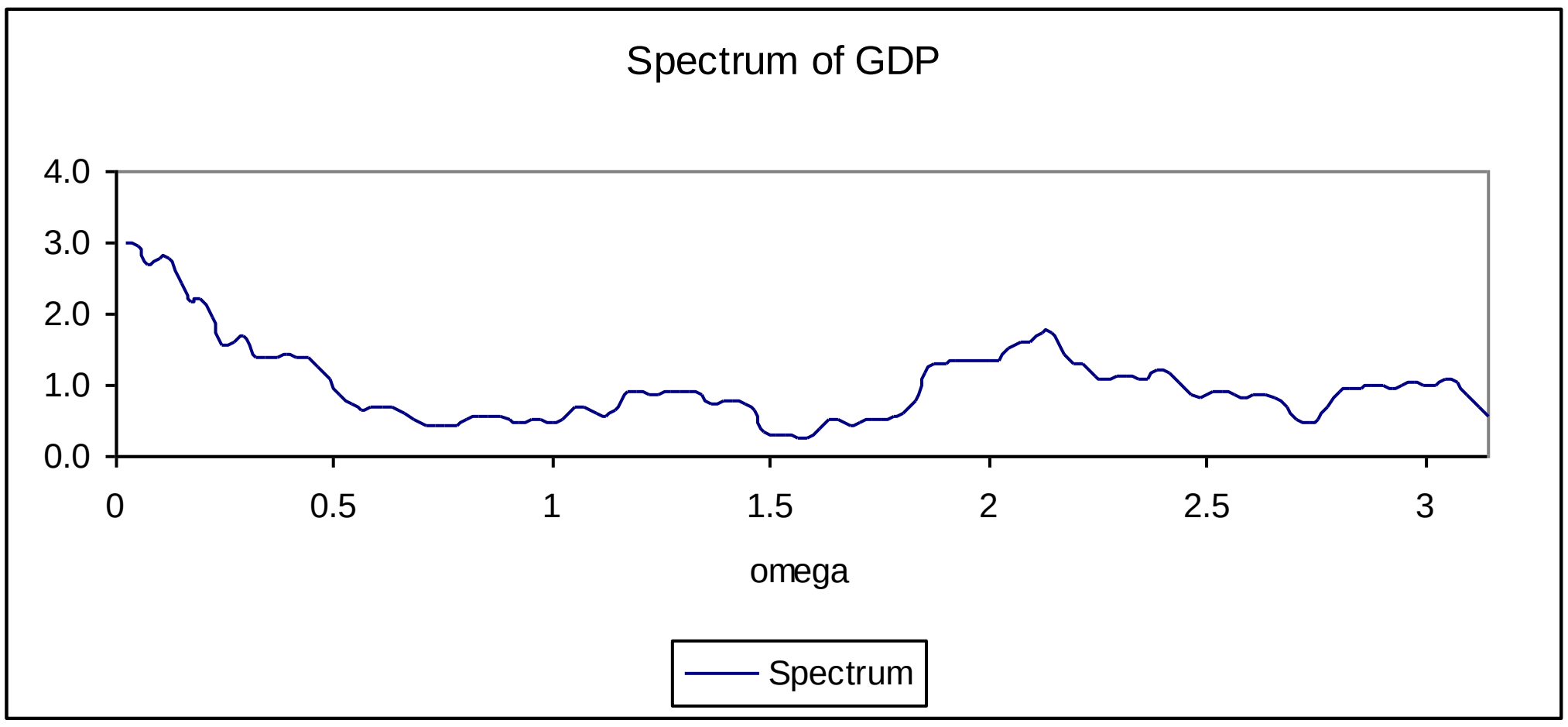

**Figure 13. Busines Cycle Patterns: GDP**